\documentclass[preprint,12pt]{elsarticle}

\usepackage{amssymb}

\usepackage{amsmath,amsfonts}
\usepackage{algorithmic}
\usepackage{algorithm}
\usepackage{array}
\usepackage[caption=false,font=normalsize,labelfont=sf,textfont=sf]{subfig}
\usepackage{textcomp}
\usepackage{stfloats}
\usepackage{url}
\usepackage{verbatim}
\usepackage{graphicx}
\usepackage{multirow}
\usepackage{threeparttable}
\usepackage{amsmath}

\usepackage[section]{placeins}
\usepackage{color}

\usepackage{float}

\journal{Computer-Aided Design}

\begin{document}

\begin{frontmatter}





\title{Topology-Aware Blending Method for Implicit Heterogeneous Porous Model Design}

\author[1]{Depeng Gao}
\author[1]{Yang Gao}
\author[1]{Yuanzhi Zhang}
\author[1]{Hongwei Lin\corref{cor1}}

\affiliation[1]{organization={School of Mathematical Sciences, Zhejiang University},
            addressline={No.866, Yuhangtang Rd}, 
            city={Hangzhou},
            postcode={310058}, 
            state={Zhejiang Provence},
            country={China}}

\cortext[cor1]{Corresponding author. E-mail address: hwlin@zju.edu.cn (H. Lin).}

\begin{abstract}
Porous structures are materials consisting of minuscule pores, where the microstructure morphology significantly impacts their macroscopic properties.  
    Integrating different porous structures through a blending method is indispensable to cater to diverse functional regions in heterogeneous models.
    Previous studies on blending methods for porous structures have mainly focused on controlling the shape of blending regions, yet they have fallen short in effectively addressing topological errors in blended structures.
    This paper introduces a new blending method that successfully addresses this issue.
    Initially, a novel initialization method is proposed, which includes distinct strategies for blending regions of varying complexities.
    Subsequently, we formulate the challenge of eliminating topological errors as an optimization problem based on persistent homology.
    Through iterative updates of control coefficients, this optimization problem is solved to generate a blended porous structure.
    Our approach not only avoids topological errors but also governs the shape and positioning of the blending region while remaining unchanged in the structure outside blending region.
    The experimental outcomes validate the effectiveness of our method in producing high-quality blended porous structures. 
    Furthermore, these results highlight potential applications of our blending method in biomimetics and the design of high-stiffness mechanical heterogeneous models.

\end{abstract}


\begin{highlights}
    \item[1.] Introduction of a novel blending method for implicit porous structures.
    \item[2.] Topological errors in the blended porous structure are avoided.
    \item[3.] Ability to freely control the shape of the blending region.
    \item[4.] Preservation of the porous structures outside the blending region.
\end{highlights}


\end{frontmatter}


\section{Introduction}
\label{sec: Introduction}
Porous structures are complex network structures composed of interconnected pores.
    Heterogeneous porous structures with varying geometric shapes in different spatial locations are common in nature.
    For instance, bone structures encompass external cortical bone with low porosity for structural reinforcement and internal cancellous bone with high porosity facilitating metabolic substance exchange~\cite{chen2020porous}.
    As a result, heterogeneous porous structures find extensive applications in bone tissue engineering and biological research.
    The morphology of porous structures significantly influences their performance. 
    Introducing diverse morphologies in different positions to create heterogeneous porous structures enables achieving superior performance compared to single-morphology structures, leading to broad applications in high-strength model design~\cite{shi2021design,feng2022stiffness,xu2023topology}. 
    Utilizing blending methods to ensure smooth transitions between different porous structures becomes essential for manufacturing and designing high-quality porous structures.

When dealing with complex-shaped porous structures, implicit porous structures such as Triply periodic minimal surfaces (TPMS)~\cite{yan2019strong}, and implicit B-splines~\cite{gao2024periodic} are commonly employed.
    Such implicit porous can be represented as a set 
\begin{equation}
    \{(x,y,z)~|~\varphi(x,y,z)\leq c(x,y,z)\}, 
\end{equation} 
where $\varphi$ represents a function field describing the porous, and $c$ is a controlling function.
    This study focuses on discussing blending techniques for implicit porous structures, for other representations, such as Volumetric representation, please refer to the relevant literature~\cite{hong2021conformal}.
    Numerous studies in Functional representation (F-rep) modeling have explored blending methods for implicit models~\cite{gourmel2013gradient,pasko1995function}, primarily concentrating on the surfaces of the models, that is, the zero iso-surfaces.
    Consequently, various blending functions have been proposed to achieve smooth and undistorted blended surfaces.
    However, existing blending methods for porous structures primarily emphasize the quality of the solid structures instead of surfaces. 
    Consequently, these techniques do not comprehensively address potential \textit{topological errors} during solid blending, that is, isolated connected components and isolated holes, potentially generating blended porous structures unsuitable for additive manufacturing applications~\cite{xu2023topology}.

Yang et al.~\cite{yang2014multi} and Yoo et al.~\cite{yoo2015advanced} introduced blending methods for controlled blending regions to achieve the blending of solid porous structures.
    However, these approaches neglect possible topological errors arising from the blending process.
    To address issues like discontinuity that emerge during blending processes, Ren et al.~\cite{ren2021transition} proposed an enhanced blending method.
    Although this improved method reduces the occurrence of topological errors, it cannot theoretically avoid these errors. 

In summary, on the one hand, blending techniques for implicit surfaces are not directly applicable to solid porous structures. 
    On the other hand, current blending methods for porous structures fail to satisfy the following criteria simultaneously: (1) Controllable blending region where engineers can specify the shape for structural control; (2) Invariance outside the blending region, ensuring deformations within blending regions do not affect external structures; (3) Absence of topological errors, generating a continuous blending structure without isolated components or holes hindering manufacturing.

In this study, we introduce a novel porous structure blending method to fulfill the aforementioned requirements.
    Initially, a blending function is initialized based on the distance to the predefined blending region boundary.
    The elimination of topological errors is then formulated as an optimization problem employing persistent homology and solved through gradient-based optimization.
    To summarize, the main contributions are as follows: 
\begin{itemize}
    \item[1.] Introduction of a novel blending method for implicit porous structures.
    \item[2.] Topological errors in the blended porous structure are avoided.
    \item[3.] Ability to freely control the shape of the blending region.
    \item[4.] Preservation of the porous structures outside the blending region.
\end{itemize}

The remainder of this study is organized as follows. 
    Subsection~\ref{subsec: Related work} provides a review of blending methods for implicit porous structures.
    Section~\ref{sec: Preliminaries} covers fundamental aspects of B-spline functions, implicit porous structures, and persistent homology.
    The proposed blending method is introduced in Section~\ref{sec: Geometric blending of porous structures}.
    Section~\ref{sec: Experiments and discussions} presents experimental results and comparisons of the proposed blending method with other techniques.
    Section~\ref{sec: Applications of the blending scheme} showcases the potential applications of the proposed blending method.
    The study is concluded in Section~\ref{sec: Conclusion}.

\subsection{Related work}
\label{subsec: Related work}
Given two porous structures, $\varphi_1 \leq c_1$, and $\varphi_2 \leq c_2$, the objective of the blending method is to find a new function field, denoted as $\varphi$, that merges these two structures into a blended porous structure $\varphi \leq 0$.
    Typically, the blended porous structure is defined as:
\begin{equation}
    \varphi = (1-\mu)(\varphi_1-c_1)+\mu(\varphi_2-c_2),
\end{equation}
where $\mu$ represents the weight function. 
    The blended porous structure achieves a continuous and smooth transition in the blending region by appropriately choosing the weight function.
    By setting weight function $\mu$ to be constantly $0$ or $1$ outside the blending region, the porous structure outside the blending region is unchanged after blending.
    In summary, the effectiveness of different blending methods for porous structures primarily relies on the selection of the weight function $\mu$. 

Yoo et al.~\cite{yoo2012heterogeneous} defined the weight function values at the vertices of the hexahedral mesh and constructed the weight function utilizing shape functions to achieve blending of TPMSs.
    The limitation of this method lies in its dependence on hexahedral mesh models, which are often challenging to obtain.
    Subsequent techniques directly defined the weight function in space to overcome this limitation of dependence on hexahedral mesh models.

Yang et al.~\cite{yang2014multi} introduced a weight function based on the Sigmoid Function (SF) to control the position of the blending region by controlling the gradient. 
    However, this method requires an implicit expression, $G(x,y,z)=0$, for the central surface of the blending region, which might be challenging to obtain for intricate blending regions. 
    To address this issue, they introduced an alternative weight function represented by the Gaussian Radial Basis Function (GRBF) to handle arbitrarily complex blending regions~\cite{yang2014multi}.
    However, the GRBF method lacks precise control over the blending region. 
    Considering the lack of intuitiveness in controlling the blending region through gradients as proposed by Yang et al., Yoo et al.~\cite{yoo2015advanced} adopted the Beta Growth (BG) function from plant simulation modeling as the weight function. 
    The parameters of the BG function can directly specify the start, end, and center of the blending region, offering a more intuitive approach to controlling the blending region.

To address issues encountered in topology optimization (TO) of heterogeneous porous models, Ozdemir et al.~\cite{ozdemir2023novel} pointed out that the SF method is not suitable for blending multiple types of TPMSs.
    Consequently, they introduced a Discrete Reconstruction (DR) method aimed at resolving this limitation.
    However, the applicability of this method is restricted to models with cubic shapes and does not extend to more complex geometries.

Inspired by biomimicry, Zhang et al.~\cite{zhang2023regulated} derived a blending method similar to the GRBF technique from the perspective of biological signal transmission.

Although the aforementioned methods have advanced the control over the blending region's shape, they often neglect the quality of the resulting blended structure.

To ensure the constant mean curvature properties of TPMSs,
Li et al.~\cite{li2021simple} proposed an optimization method for blended structures using a modified Allen-Cahn equation. 
    This method achieves a smooth transitional structure without stress concentration. 
    Ren et al.~\cite{ren2021transition} introduced an improved weight function to address potential discontinuities in the SF blending method.     
    However, this new weight function involves numerous hyperparameters and does not address the elimination of isolated holes.

In conclusion, current blending methods for porous structures primarily focus on shaping the blending regions while overlooking the quality of the resulting blended structure.
    Although some studies have emphasized the smoothness and connectivity of blended structures, there is limited research on eliminating higher-dimensional topological errors, such as isolated holes. 
    Therefore, this study proposes a novel blending method that not only ensures controllable blending regions but also theoretically eliminates topological errors in the blended structure to address the aforementioned issues. 

\section{Preliminaries}
\label{sec: Preliminaries}

\subsection{B-spline functions}
\label{subsec: B-spline functions}
Let $\mathbf{U}=\{u_0,\ldots,u_{m-1}\}$ be a nondecreasing sequence of real numbers, which is named as knot vector. 
    The $i$-th B-spline basis $N_{i,p}(u)$ with a degree of $p$ is defined based on the knot vector $\mathbf{U}$.
    A B-spline function $C(u)$ with degree of $p$ is defined as follows:
\begin{equation}
    C(u) = \sum_{i=0}^{M-1}N_{i,p}(u)C_i ~~u_0\leq u\leq u_{m-1},
\end{equation}
where $C_i\in\mathbb{R}$ is the $i$-th control coefficient~\cite{piegl2012nurbs}.

The support domain of the basis $N_{i,p}(u)$ is the interval $[u_i,u_{i+p+1})$.
    Consequently, adjustment of the control coefficient $C_i$ only affects the interval $[u_i,u_{i+p+1})$.
    
A trivariate B-spline function is a tensor product of the univariate B-spline functions. 
    A trivariate B-spline function with degree of $(p_u,p_v,p_w)$ is defined as:
\begin{equation}
    C(u,v,w) = \sum_{i=0}^{M_u-1}\sum_{j=0}^{M_v-1}\sum_{k=0}^{M_w-1}N_{i,p_u}(u)N_{j,p_v}(v)N_{k,p_w}(w)C_{ijk},
\end{equation}
where $C_{ijk}\in\mathbb{R}$ is the control coefficient, and, $M_u$, $M_v$ and $M_w$ are the number of control coefficients in each direction.

\subsection{Implicit porous structures}
\label{subsec: Design of implicit porous model}
Due to the intricate topology and geometry of porous structures, representing them using the commonly utilized Boundary representation in CAD software is challenging. 
    Therefore, in recent years, numerous studies have opted for implicit functions to represent and design porous structures~\cite{yan2019strong,feng2019efficient}.     
    Given a function field $\varphi(x,y,z)$ defined in $\mathbb{R}^3$, a porous structure $\Phi$ can be defined as:
\begin{equation}
    \Phi = \{(x,y,z)~|~\varphi(x,y,z)\leq c(x,y,z)\},
\end{equation}
where $c(x,y,z)$ represents the threshold distribution field (TDF)~\cite{hu2021heterogeneous}. 

The function $\varphi(x,y,z)$ is commonly utilized to determine the type of porous structure. 
    In implicit porous structure design, TPMSs~\cite{hu2021heterogeneous,yoo2011porous} or implicit B-splines~\cite{hong2023implicit,gao2022connectivity,gao2024periodic} are frequently employed as the function field $\varphi(x,y,z)$. 
    Once the type of porous structure is determined, the TDF $c(u,v,w)$ becomes crucial for controlling the porous structure. 
    Several studies have demonstrated that manipulating the TDF can regulate crucial properties of implicit porous structures such as relative density, stiffness, and anisotropy, among others~\cite{hu2021heterogeneous,hu2023isogeometric,feng2021isotropic}. 

Furthermore, based on different set definitions, implicit porous structures can be categorized into three types:
\begin{itemize}
    \item[1.]  \textbf{Pore} type: $\Phi = \{(x,y,z)~|~\varphi(x,y,z)\geq c(x,y,z)\}$
    \item[2.]  \textbf{Rod} type: $\Phi = \{(x,y,z)~|~\varphi(x,y,z)\leq c(x,y,z)\}$
    \item[3.]  \textbf{Sheet} type: $\Phi = \{(x,y,z)~|~ c_1(x,y,z)\leq \varphi(x,y,z)\leq c_2(x,y,z)\}$,
\end{itemize}
where $c(x,y,z)$, $c_1(x,y,z)$, and $c_2(x,y,z)$ represent TDFs.

In the manufacture process of porous structures, isolated connected components and isolated holes cannot be manufactured. 
    They are named as zero-dimensional and two-dimensional topological errors, respectively.
    In this study, we focus on eliminating these topological errors.

\subsection{Persistent homology}
\label{subsec: Persistent homology}
Persistent homology serves as a crucial method in Topological Data Analysis (TDA) for capturing the topological information of data sets filtered by real-valued functions. 
    This study employs implicit functions to represent porous structures, naturally inducing a sub-level filtration that enables the calculation and optimization of the topological features using persistent homology.

In Euclidean space, an elementary interval is defined as $I = [l, l+1]$, where $l \in \mathbb{Z}$. 
    A $k$-cube $\kappa_k$ is the Cartesian product of $k$ elementary intervals.  
    A cubical complex $\mathcal{K}$ is a collection of $k$-cubes, where every face of a cube should be contained in $\mathcal{K}$.
    The union of all cubes in $\mathcal{K}$ forms the underlying space $\mathcal{S}$.

\begin{figure}[h]
\centering
    \includegraphics[width = 0.8\textwidth]{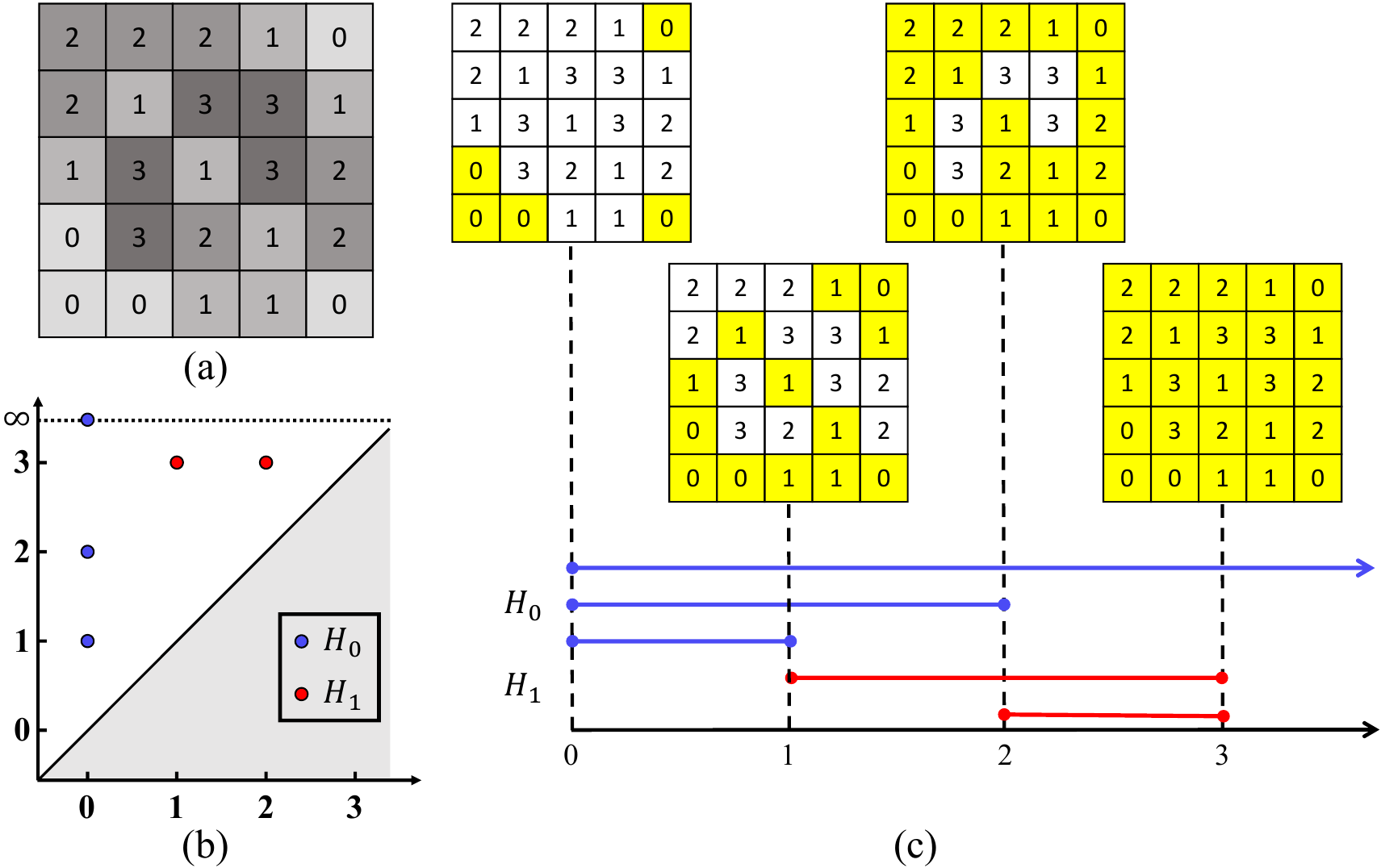}    
        \caption{Illustration of persistent homology. (a) A cubical complex $\mathcal{K}$ with the value of $f(\sigma)$ on it for each $\sigma \in \mathcal{K}$. (b) Persistent pairs $(b, d)$ are shown in the form of a persistent diagram. (c) Persistent pairs $(b, d)$ are shown in the form of persistent barcodes, and cubical complexes are marked in yellow. Blue segments and red segments correspond to the $0$-dimensional and $1$-dimensional topological features, respectively. At time $0$, there are three connected components in the underlying space. At the time $1$, one of the connected components disappears and a loop appears. At time $2$, another connected component is merged and the second loop is formed. At time $3$, both of the two loops disappear. Because there is always a connected component in the underlying space, we can see a blue ray that extends to infinity and a point (0,$\infty$) in the persistent diagram.}
        \label{fig: persistent_homology}
\end{figure}    

The $d$-th homology group over the complex $\mathcal{K}$ is denoted as $H_d(\mathcal{K})$, encoding the $d$-dimensional topological features of $\mathcal{K}$. 
    The rank of $H_d(\mathcal{K})$ is termed the $d$-Betti number $\beta_d(\mathcal{K})$, indicating the number of homological classes in dimension $d$. 
    Under this definition, $\beta_0(\mathcal{K}), \beta_1(\mathcal{K})$ and $\beta_2(\mathcal{K})$ represent the counts of connected components, loops and voids within $\mathcal{S}$, respectively.

Let $f : \mathcal{S} \rightarrow \mathbb{R}$ be a real-valued function. 
    The value of $\tau \in \mathcal{K}$ is defined as the maximum of all points in $\tau$, denoted as $f(\tau) = \max_{p \in \tau} f(p).$
    Subsequently, the induced cubical complex $\mathcal{K}_f^t := \left\{\tau \in \mathcal{K} | f(\tau) \le t \right\}$ is formed, where $t\in \mathbb{R}$.  
    The sub-level filtration $\mathcal{K}_f$ is defined as a nested sequence of cubical complexes $\mathcal{K}_f^{t_0} \subseteq \mathcal{K}_f^{t_1} \subseteq \cdots \subseteq \mathcal{K}_f^{t_m}$. 
    As the parameter $t_i$ increases, the ``birth" and ``death" of topological features in different dimensions can be captured.
    The persistent pairs $(b, d)$ can be utilized to represent these topological features and plotted into a persistent diagram (see Figure~\ref{fig: persistent_homology}(b)) or a persistent barcode (see Figure~\ref{fig: persistent_homology}(c)).

\section{Geometric blending of porous structures}
\label{sec: Geometric blending of porous structures}
Given porous structures with varying shapes, porosity, and pore sizes in adjacent regions, the aim of the blending problem is to establish a plausible transition within these regions. 
    This section begins by discussing the blending scheme for two porous structures. 
    The scheme involves initializing a blended porous structure (refer to Subsection~\ref{subsec: 3.1 Initialization of blending function}) and subsequently optimizing this structure (refer to Subsection~\ref{subsec: 3.2 Optimization of blending function}). 
    Furthermore, this blending problem firstly focused on two porous structures that can naturally progress to encompass multiple porous structures (refer to Subsection~\ref{subsec: 3.3 Blending scheme for more than two porous structures}).

\begin{figure}[h]
\centering
    \includegraphics[width = 0.98\textwidth]{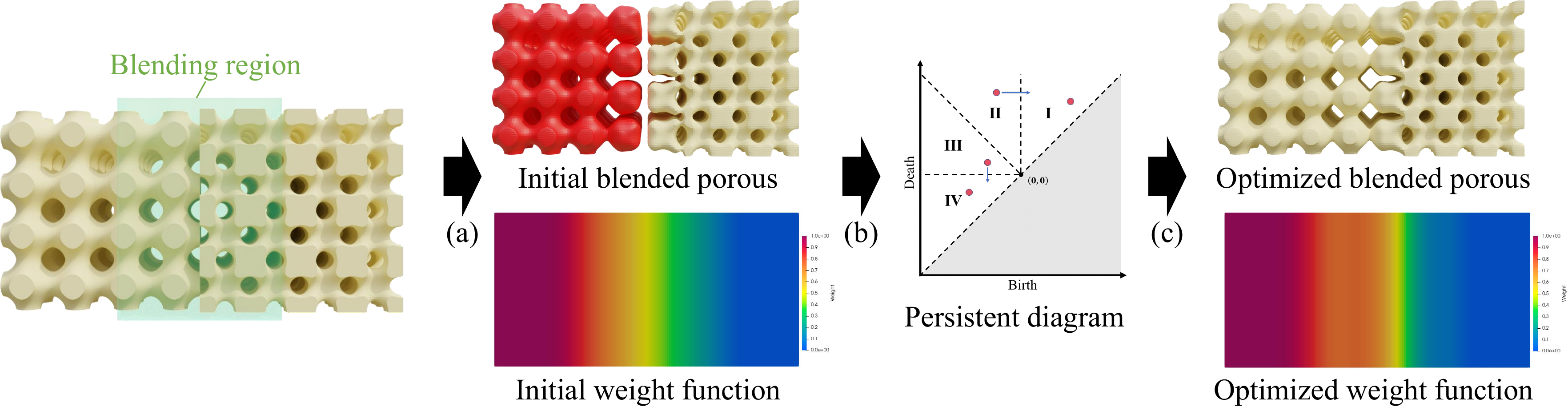}    
        \caption{Illustration of the pipeline of the proposed method. (a) The weight function is initialized, followed by the generation of the initial blended porous structure. The isolated connected component in the initial blended porous is marked in red. (b) The persistence diagram which records the topological information of the blended porous structure is calculated. (c) The weight function is iteratively optimized and guided by the persistence diagram, and correspondingly, the blended porous structure is also updated. }
        \label{fig: Flowchart}
\end{figure}

The pipeline of the introduced blending scheme is depicted in Figure~\ref{fig: Flowchart}. 
    Initially, an initial blending function is obtained to generate the initial blended structure. 
    Following this, the objective function is formulated by computing the representatives of the homology class of the initial blended porous structure. 
    Ultimately, the blending function undergoes iterative optimization, resulting in a blended porous structure without topological errors.

\subsection{Initialization of blending function}
\label{subsec: 3.1 Initialization of blending function}
A robust initialization can significantly impact the optimization process, leading to superior solutions. 
    Therefore, this section discusses the initialization of the blending function. 
    In cases where the adjacent boundaries of two porous structures are simple, a one-dimensional blending function can be employed to create the initial blended porous structure (refer to Subsection~\ref{subsec: 3.1.1 One-dimensional blending situation}). 
    However, when dealing with complex adjacent boundaries, a three-dimensional blending function becomes necessary to achieve the blend (refer to Subsection~\ref{subsec: 3.1.2 Three-dimensional blending situation}). 

\subsubsection{One-dimensional initialization}
\label{subsec: 3.1.1 One-dimensional blending situation}
Although porous structures are typically defined in three-dimensional Euclidean space, the practice of blending along one direction is prevalent in various applications~\cite{vijayavenkataraman2018triply,xu2023topology,ren2021transition}.
    Employing one-dimensional blending functions in these scenarios proves to be more efficient in terms of memory consumption and computational time.
    The most straightforward scenario involves blending two porous structures in one direction within the cartesian coordinate system. 
    The blending along specific boundaries, such as cylindrical or spherical surfaces, can be simplified into one-dimensional scenarios using the cylindrical and spherical coordinate systems, respectively.
    These scenarios are discussed in this subsection.

\begin{figure*}[h]
\centering
    \includegraphics[width = 0.98\textwidth]{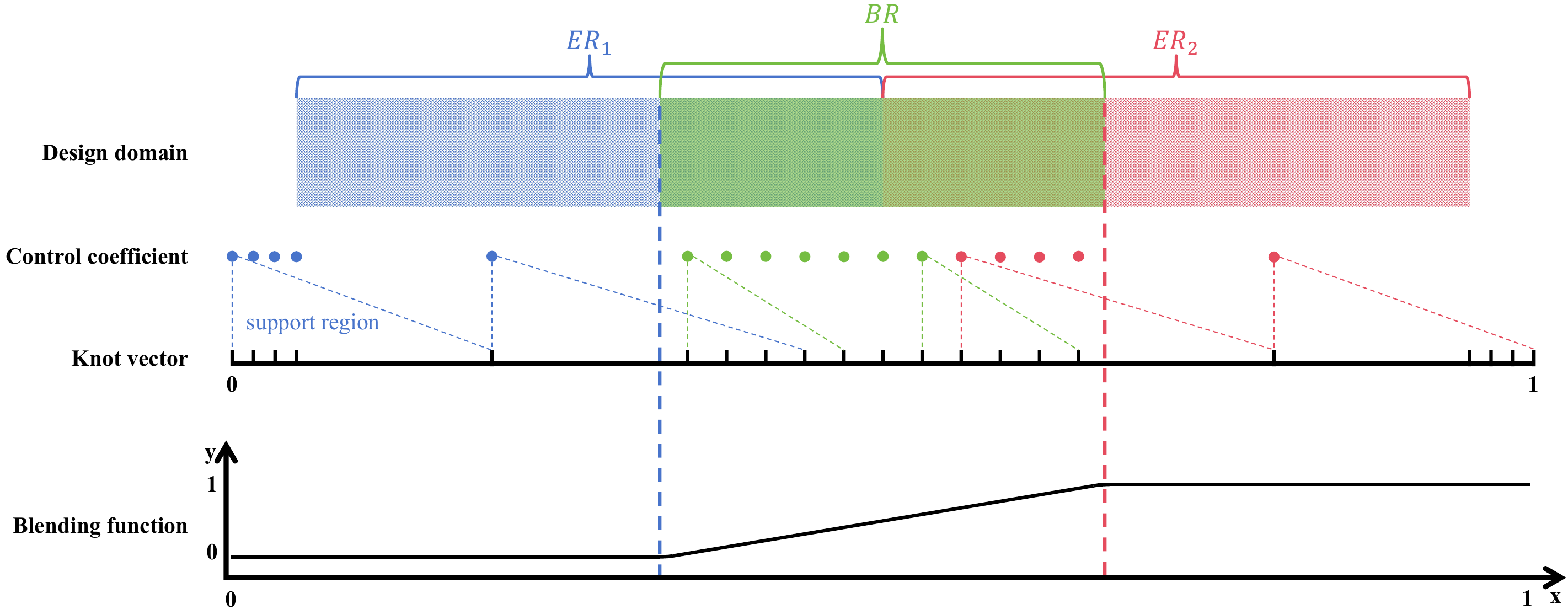}    
        \caption{Illustration of one-dimensional initial blending function. For ease of understanding, the BR (blending region) and ER (existing region) are drawn as two-dimensional regions. Two porous structures in $ER_1$ and $ER_2$ are blended along the $x$-direction in $BR$. To generate a blending function like at the bottom, control coefficients marked in blue and red are set to 0 and 1, respectively. Additionally, control coefficients marked in green are uniformly assigned values from 0 to 1 from left to right.}
        \label{fig: Illustration of one-dimensional initial blending function}
\end{figure*}

\textbf{Cartesian coordinate system. }Assuming that two porous structures represented as sets $\Phi_1 = \{\varphi_1\leq c_1\}$ and $\Phi_2=\{\varphi_2 \leq c_2\}$ situated in distinct existing regions, denoted as $ER_1$ and $ER_2$ (refer to Figure~\ref{fig: Illustration of one-dimensional initial blending function}). 
    The total existing region $TER$ is the union of these two existing regions. 
    In the proposed scheme, users should assign the location of the transition structure at the blending region, referred to as $BR$ (blending region).
    The objective is to achieve a blended porous structure that retains the internal structures of $ER_1 \setminus BR$ and $ER_2 \setminus BR$ unchanged, featuring a smooth transition structure within $BR$.
    The ideal blending function $\omega(x,y,z)$ is defined as follows:
\begin{equation}
    \omega(x,y,z) = 
    \begin{cases}
        0 & (x, y, z)\in ER_1 \setminus BR \\
        1 & (x, y, z)\in ER_2 \setminus BR \\
        \alpha & (x, y, z)\in BR,~\alpha \in [0,1],
    \end{cases}
\end{equation}
   
To leverage the accurate fitting and local support properties of B-splines, B-spline functions are chosen as the blending function to enable precise control of the blending region and optimization of topological features. 
    Given that the parametric domain of B-splines is typically defined in $[0,1]$, $TER$ is scaled to fit within $[0,1]\times[0,1]\times[0,1]$, reverting to its original size post-generation of the final blended porous structure.
  
As illustrated in Figure~\ref{fig: Illustration of one-dimensional initial blending function}, the two-dimensional regions are used for clarity, and the blending of two porous structures occurs along the $x$-direction. 
    Subsequently, the initial blending function is defined as:
\begin{equation}
    \omega(x,y,z) = \omega(x) = \sum_{i=0}^{n-1}B_i(x)C_i.
\end{equation} 
    To maintain the porous structures in $ER_1 \setminus BR$, the control coefficients (depicted as blue dots in Figure~\ref{fig: Illustration of one-dimensional initial blending function}) whose support domains intersect with $ER_1 \setminus BR$ are set to $0$.
    Similarly, control coefficients influential on $ER_2 \setminus BR$ are assigned the value of $1$ (as indicated by red dots in Figure~\ref{fig: Illustration of one-dimensional initial blending function}). 
    Control coefficients whose support domains are contained in $BR$ are uniformly assigned values from 0 to 1 from left to right.
    Finally, the blended porous structure $\Phi_{ini}$ and its corresponding function $\varphi_{ini}$ are defined as:
\begin{equation}
    \left\{
    \begin{aligned}
        &\Phi_{ini}=\{(x,y,z)~|~\varphi_{ini}(x,y,z)\leq 0\}\\
        &\varphi_{ini}(x,y,z) = (1-\omega)(\varphi_1-c_1)+\omega(\varphi_2-c_2).
    \end{aligned}
    \right.
    \label{eq: initial blended porous structure}
\end{equation}

\begin{figure}[h]
\centering
    \includegraphics[width = 0.78\textwidth]{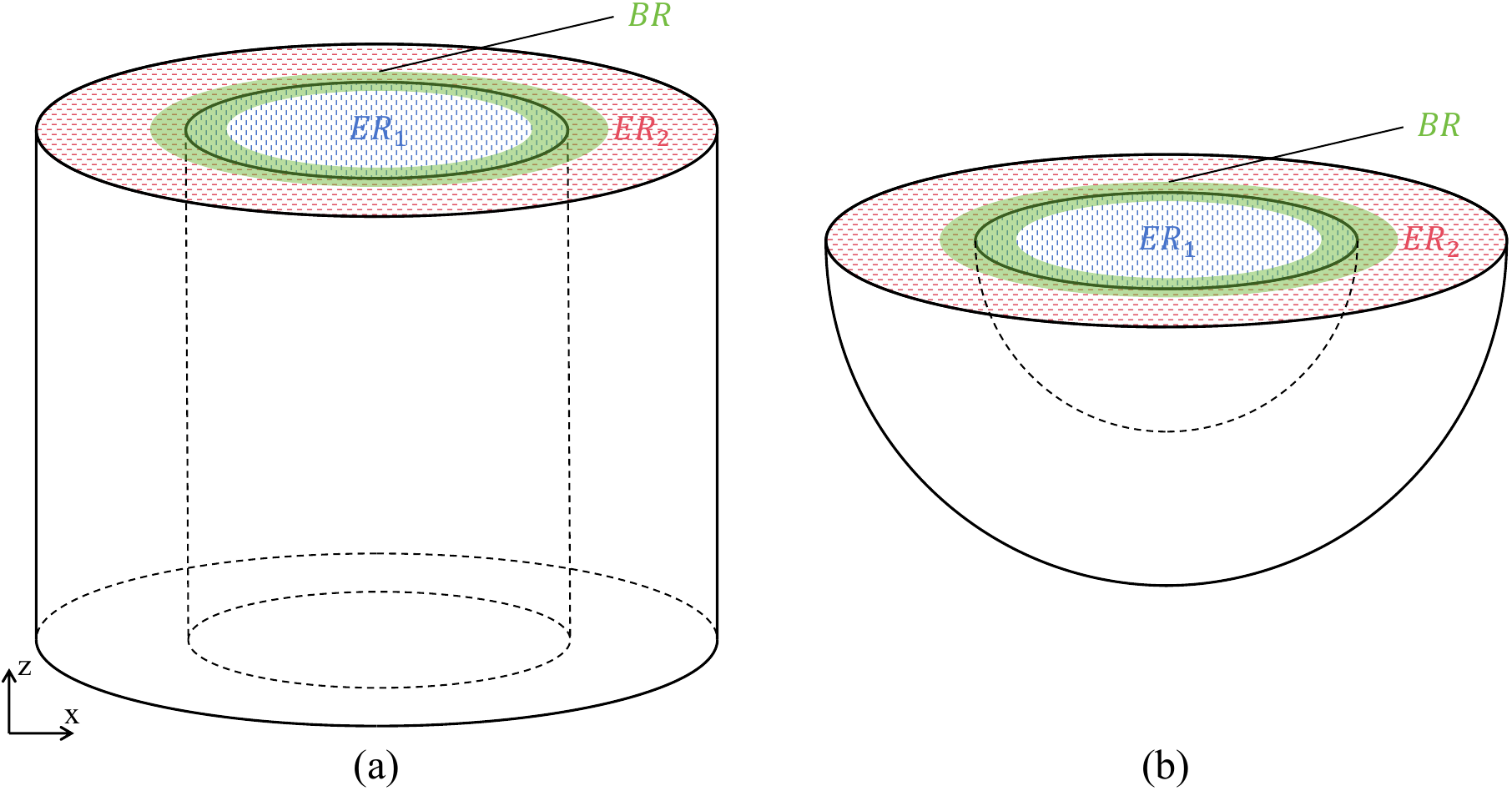}    
        \caption{Two porous structures defined in regions $ER_1$ and $ER_2$ are blended in annular region $BR$. At this point, the weight function can be viewed as a one-dimensional function of radius $r$.}
        \label{fig: Cylinder blending}
\end{figure} 

\textbf{Cylindrical coordinate system. }As illustrated in Figure~\ref{fig: Cylinder blending}(a), two porous structures are blended radially along the cylinder. 
The adjacent boundary of $ER_1$ and $ER_2$ is a three-dimensional cylindrical surface:
\begin{equation}
f(x,y,z) = x^2+y^2-R^2=0.
\end{equation} 
In this scenario, the blending region is modeled as a concentric cylindrical structure.
    The blending function $w(x,y,z)$ is expressed as a function of the distance $r$ to the central axis of the cylinder:
\begin{equation}
\begin{cases}
    w(x,y,z) = B(r) = \sum_{i=0}^{n}B_i(r)R_i \\
    r = \sqrt{x^2 + y^2},
\end{cases} 
\label{eq: cylinder blending}
\end{equation}
where $R_i$ is the control coefficients and $B_i(r)$ is the $i$-th B-spline basis function.
    Subsequently, the blending function $B(r)$ can be initialized using the method illustrated in Figure~\ref{fig: Illustration of one-dimensional initial blending function} as in the cartesian coordinate system.

\textbf{Spherical coordinate system. }As illustrated in Figure~\ref{fig: Cylinder blending}(b), the adjacent boundary of $ER_1$ and $ER_2$ is a three-dimensional spherical surface:
\begin{equation}
f(x,y,z) = x^2+y^2+z^2-R^2=0.
\end{equation} 
In this scenario, the blending function $w(x,y,z)$ is expressed as a function of the distance $r$ to the midpoint of the sphere:
\begin{equation}
\begin{cases}
    w(x,y,z) = B(r) = \sum_{i=0}^{n}B_i(r)R_i \\
    r = \sqrt{x^2 + y^2 + z^2}.
\end{cases} 
\label{eq: shpere blending}
\end{equation}
    Subsequently, the blending function $B(r)$ can be initialized using the method illustrated in Figure~\ref{fig: Illustration of one-dimensional initial blending function} as in the cartesian coordinate system.

\subsubsection{Three-dimensional initialization}
\label{subsec: 3.1.2 Three-dimensional blending situation}
When the adjacent boundary of two porous structures is a complex surface, the straightforward definition of the blending function as a one-dimensional function becomes impractical.
    In this scenario, a three-dimensional blending function is necessary to achieve an initial blended porous structure.

\begin{figure}[h]
\centering
    \includegraphics[width = 0.68\textwidth]{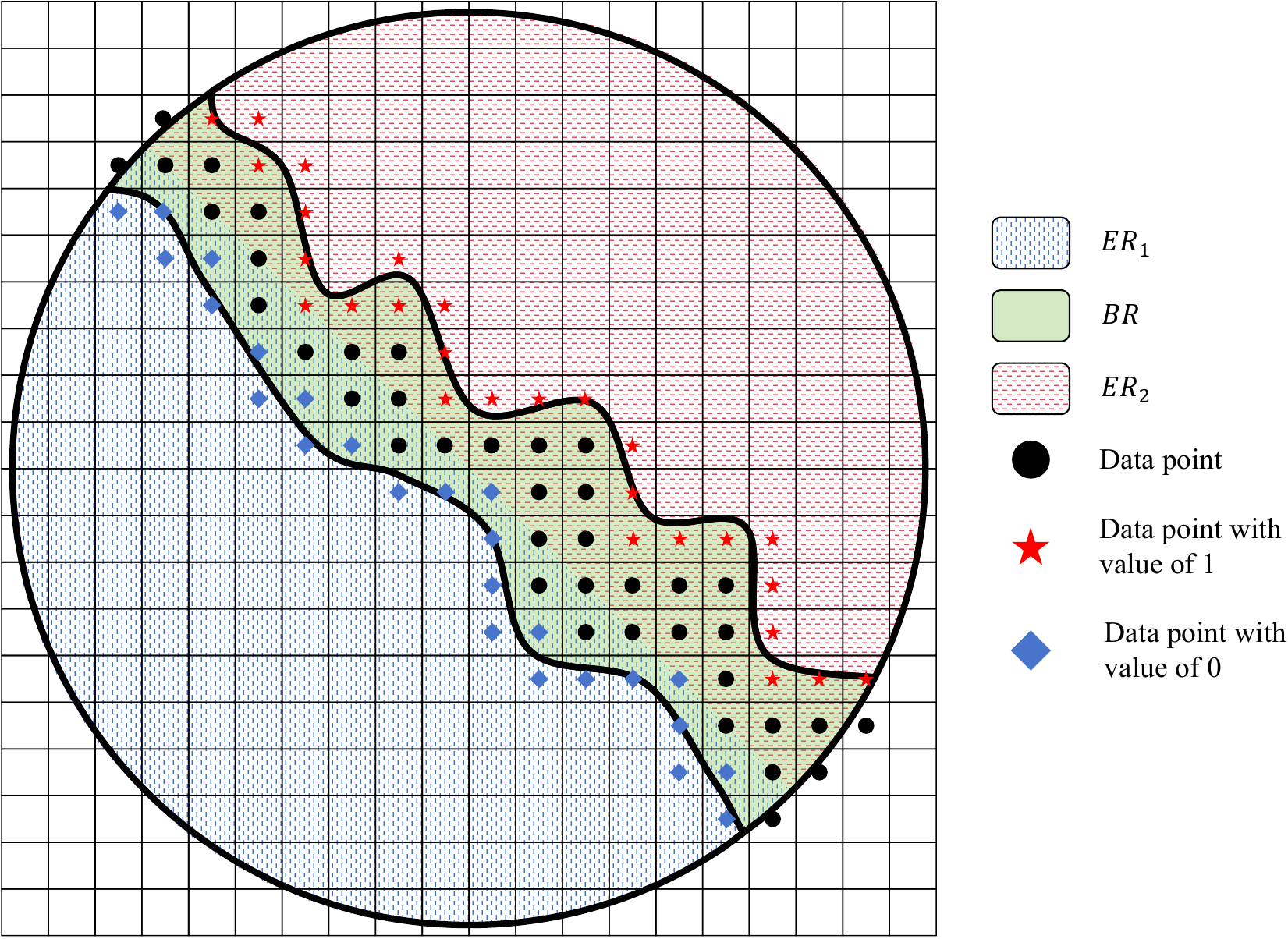}    
        \caption{Illustration of three-dimensional initial blending function. For ease of understanding, the BR (blending region) and ER (existing region) are drawn as two-dimensional regions. }
        \label{fig: Illustration of three-dimensional initial blending function}
\end{figure} 

As illustrated in Figure~\ref{fig: Illustration of three-dimensional initial blending function}, the two-dimensional regions are used for ease of understanding, and two porous structures are blended within the user-specified blending region.
    The blending function is established through a three-dimensional B-spline function given by: 
\begin{equation}
    \begin{aligned}
        &\omega(x,y,z) = \sum_{i=0}^{n_u-1}\sum_{j=0}^{n_v-1}\sum_{k=0}^{n_w-1}R_{ijk}(u,v,w)C_{ijk} \\
        &R_{ijk}(u,v,w) = B_i(u)B_j(v)B_k(w),
    \end{aligned}
\end{equation}
where $R_{ijk}(u,v,w)$ is the blending basis function.
    Subsequently, the control coefficient $C_{ijk}$ are initialized as outlined in Subsection~\ref{subsec: 3.1.1 One-dimensional blending situation}:
\begin{equation}
    C_{ijk} = 
    \begin{cases}
        0 & \exists \gamma \in ER_1 \text{ s.t. } R_{ijk}(\gamma)\neq 0 \\
        1 & \exists \gamma \in ER_2 \text{ s.t. } R_{ijk}(\gamma)\neq 0.
    \end{cases}
    \label{eq: three-dimensional initialization}
\end{equation}
    It is noteworthy that if a basis function's support region intersects both $ER_1$ and $ER_2$ simultaneously, the knot insertion algorithm~\cite{piegl2012nurbs} must be employed to refine the nodal lines in the blending region.

\begin{figure}[H]
\centering
\resizebox{0.8\textwidth}{!}{%
\begin{minipage}{\textwidth}
\begin{algorithm}[H]
    \renewcommand{\algorithmicrequire}{\textbf{Input:}}
	\renewcommand{\algorithmicensure}{\textbf{Output:}}
    \caption{Generate Initial Blending Function}
    \label{algorithm: Generate Initial Blending Function}
    \begin{algorithmic}[1]
\STATE Discretize the space into a uniform grid of cells
\FOR{each \textit{cell} in the grid}
    \IF{\textit{cell} intersects with $ER_1$ and the boundary of $BR$}
        \STATE Add center of \textit{cell} to $S_{boundary}^0$.
    \ELSIF{\textit{cell} intersects with $ER_2$ and the boundary of $BR$}
        \STATE Add center of \textit{cell} to $S_{boundary}^1$.
    \ELSIF{\textit{cell} is in $BR$ but not intersected with its boundary}
        \STATE Add center of \textit{cell} to $S_{inner}$ as an interior point
    \ENDIF
\ENDFOR
\FORALL{points $p \in S_{boundary}^i$, $i=0$ or $1$}
    \STATE $v_p \gets i$
\ENDFOR
\STATE Construct KD-trees from points in $S_{boundary}^0$ and $S_{boundary}^1$
\FORALL{points $p \in S_{inner}$}
    \STATE $v_p \gets$ $dis(p, S_{boundary}^0) /(dis(p, S_{boundary}^0)+dis(p, S_{boundary}^1)) $
\ENDFOR
\STATE Construct data points set $\{p,v_p\}$, where $p\in S_{inner} \cup S_{boundary}^0 \cup S_{boundary}^1$ 
\STATE Use Local-LSPIA algorithm to fit the data points $\{p,v_p\}$ with a B-spline function $\omega(x, y, z)$
\ENSURE $\omega(x, y, z)$
    \end{algorithmic}
\end{algorithm}
\end{minipage}%
}
\end{figure}
    
The blending function should be 0 in $ER_1 \setminus BR$ and 1 in $ER_2 \setminus BR$, transitioning gradually from 0 to 1 within $BR$.
    The fitting method is used to create the blending function, with the algorithm flowchart presented in Algorithm~\ref{algorithm: Generate Initial Blending Function}.
    The space is initially discretized into a uniform grid.
    If a grid cell intersects both $ER_1$ and the boundary of $BR$ simultaneously, the center is considered a 0-boundary point (as indicated by blue rhombuses in Figure~\ref{fig: Illustration of three-dimensional initial blending function}). 
    Similarly, when a grid cell intersects both $ER_2$ and the boundary of $BR$ at the same time, its center is identified as a 1-boundary point (illustrated by red stars in Figure~\ref{fig: Illustration of three-dimensional initial blending function}).
    A cell within $BR$ but not intersecting its boundary is considered an interior point (depicted as black dots in Figure~\ref{fig: Illustration of three-dimensional initial blending function}).
   
All interior points form set $S_{inner}$, while all 0-boundary and 1-boundary points form set $S_{boundary}^0$ and $S_{boundary}^1$, respectively. 
    Each point $p\in S_{boundary}^i$ corresponds to $v_p=i$ where $i=0$ or $1$, while each point $p\in S_{inner}$ is associated with a value defined as follows:
\begin{equation}
    v_p = \frac{dis(p,S_{boundary}^0)}{dis(p,S_{boundary}^0)+dis(p,S_{boundary}^1)},
\end{equation}
where $dis(p,S_{boundary}^i)$ is the Euclidean distance form $p$ to set $S_{boundary}^i$, $i=0,1$. 
    For efficiency, KD-trees are built using sets $S_{boundary}^0$ and $S_{boundary}^1$ to determine the nearest neighbor in these sets for each point in $S_{inner}$ as the distance.  

By fitting all above data points, the initial weight function can be obtained. 
    However, directly solving the fitting problem will result in the values of the weight function outside the blending region $BR$ not being constantly 0 or 1. 
    Therefore, constraints need to be imposed on the fitting problem. 
    The set $I_{fixed}$ consisting of subscripts of the control coefficients that affect region outside the blending region $BR$ is defined as follows: 
\begin{equation}
    I_{fixed} = \{(i,j,k)~|~\exists (x,y,z)\notin BR ~\text{s.t.}~R_{ijk}(x,y,z)\neq 0\}.
    \label{eq: fixed subscripts}
\end{equation}
    During the fitting process, the control coefficients corresponding to the index set $I_{fixed}$ should remain unchanged to ensure the invariance of the porous structure outside the blending region $BR$.
    Finally, a constrainted fitting problem is formulated to derive the initial blending function based on the data points $S_{boundary}^0$, $S_{boundary}^1$, $S_{inner}$, and their associated values:
\begin{equation}
\begin{aligned}
    \min_{\omega(x,y,z)} \quad &\sum_{p\in S_{boundary}^0 \cup S_{boundary}^1\cup S_{inner} }\Vert \omega(p)-v_p \Vert_2^2. \\
    \text{s.t.} \quad & \omega(x,y,z) = \sum_{i=0}^{n_u-1}\sum_{j=0}^{n_v-1}\sum_{k=0}^{n_w-1}R_{ijk}(p)C_{ijk} \\
    & C_\zeta \text{ is fixed, where } \zeta\in I_{fixed}
\end{aligned}
    \label{eq: fitting problem}
\end{equation}
    To efficiently address the complex fitting problem defined in Equation~\ref{eq: fitting problem}, the Local Progressive and Iterative Approximation for Least Squares Algorithm~\cite{local-lspia} (Local-LSPIA) is employed for iterative solutions.
    
\subsection{Optimization of blending function}
\label{subsec: 3.2 Optimization of blending function}
In Subsection~\ref{subsec: 3.1 Initialization of blending function}, the initial blended porous structure $\Phi_{ini}$ and its corresponding function $\varphi_{ini}$ are obtained utilizing the local support property of B-splines. 
    It is important to note that while employing a high-degree B-spline function $\omega$ can yield a smooth function $\varphi_{ini}$, as depicted in the second column of Figure~\ref{fig: Flowchart}, the resulting blended porous structure may be disconnected. 
    This disconnected arises from the fact that the blended structure $\Phi_{ini}$ is created by truncating the smooth function $\varphi_{ini}$ (refer to Equation~\ref{eq: initial blended porous structure}).
    Consequently, further optimization of the initial blended structure $\Phi_{ini}$ and its corresponding function $\varphi_{ini}$ is necessary to avoid topological errors such as this disconnected.
    
As explained in Section~\ref{subsec: Persistent homology}, when function $f(x,y,z)$ is employed to induce a filtration, the resulting persistent diagram captures persistent pairs (corresponding to topological features) of the set $\{(x,y,z)~|~f(x,y,z)\leq c\}$ as $c$ increases from $-\infty$ to $\infty$.
    Therefore, by substituting $f(x,y,z)$ with $\varphi_{ini}$, the topological information of the initial blended porous structure $\Phi_{ini}$ can be revealed. 
    The resulting $k$-dimensional persistent diagrams consist of the following persistent pairs:
\begin{equation}
    P_k = \{(b_{i,k},d_{i,k})\}_{i=0}^{N_k-1},
    \label{eq: persistent pairs}
\end{equation}
where $b_i^k$ and $d_i^k$ are the birth time and death time of the $i$-th persistent pair in $k$-dimensional situation. 
    
The blended porous structure $\Phi_{ini}$ is represented as $\{\varphi_{ini}(x,y,z)\leq c=0\}$ (refer to Equation~\ref{eq: initial blended porous structure}). 
    Therefore, only persistent pairs that birth before $0$ (above the line $y=0$), and die after $0$ (left of $x=0$), exist in the blended porous structure $\Phi_{ini}$. 
    Isolated holes emerge in the blended porous structure when two-dimensional persistent pairs exist, while isolated connected components arise when multiple zero-dimensional persistent pairs are present. 
    The existence of these persistent pairs compromises the manufacturability of the blended porous structure.
    Therefore, we need to remove these persistent pairs to obtain a manufacturable blended porous structure.

\begin{figure}[h]
\centering
    \includegraphics[width = 0.5\textwidth]{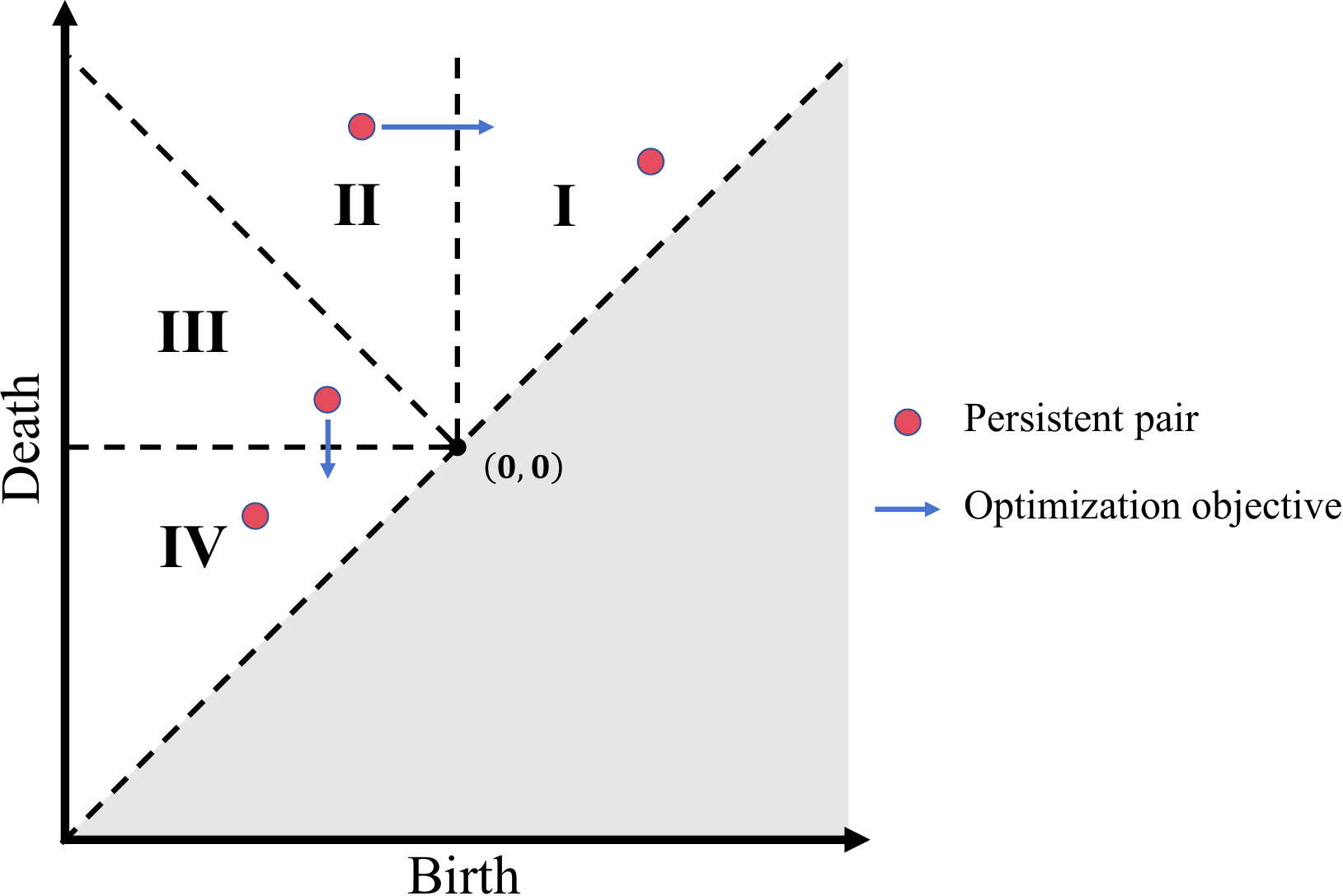}    
        \caption{A two-dimensional persistent diagram. Regions I, II, III, and IV are segmented by lines $x=0$, $y=0$, and $y=-x$. The persistent pairs in regions II and III correspond to the isolated holes within the blended porous structure. The blue arrows indicate the direction of the optimization objective.}
        \label{fig: illustration of optimization objective}
\end{figure} 

A two-dimensional persistent diagram as shown in Figure~\ref{fig: illustration of optimization objective} is utilized to illustrate the optimization objective. 
    Death times are always larger than birth times, situating all persistent pairs above the line $y=x$. 
    The diagram is segmented by the lines $y=x$, $x=0$, $y=0$, and $y=-x$ into four regions. 
    Only persistent pairs in regions II and III correspond to isolated holes in the blended porous structure. 
    To address these isolated holes, persistent pairs from regions II and III should be translocated to regions I and IV.
    This transfer of persistent pairs impacts the underlying porous structure concurrently. 
    Consequently, persistent pairs in region II are translocated to region I, and, persistent pairs in region III are translocated to region IV to avoid a great change of porous structure caused by a considerable movement of persistent pairs.
    In a zero-dimensional situation, the set $\{\varphi_{ini}(x,y,z)\leq +\infty\}$ is always one connected component, which corresponds to a persistent pair $(b,+\infty)$, $b\in \mathbb{R}$ in the zero-dimensional persistent diagram. 
    We default that this persistent pair is not in regions II and III.
    Subsequently, the optimization objective treats zero-dimensional persistent pairs the same as their two-dimensional counterparts.

The $k$-dimensional persistent pairs $P_k$ defined in Equation~\ref{eq: persistent pairs} are categorized into distinct sets based on regions:
\begin{equation}
    \left\{
    \begin{aligned}
        &P_k^{\text{II}} = \{(b_{i,k}^\text{II},d_{i,k}^\text{II})\}_{i=0}^{N_k^\text{II}-1}\\
        &P_k^{\text{III}} = \{(b_{i,k}^\text{III},d_{i,k}^\text{III})\}_{i=0}^{N_k^\text{III}-1},
    \end{aligned}
    \right.
    \label{eq: persistent pairs in different regions}
\end{equation}
where $N_k^\text{II}$ and $N_k^\text{III}$ denote the counts of persistent pairs within regions II and III, respectively.
    The repositioning of zero-dimensional and two-dimensional persistent pairs from region III to IV and region II to I, respectively, is achieved by defining the objective function as:
\begin{equation}
    L = \sum_{k=0,2}\left( \sum_{i=0}^{N_k^\text{III}-1}d_{i,k}^\text{III} -
    \sum_{i=0}^{N_k^\text{II}-1}b_{i,k}^\text{II}\right).
    \label{eq: optimization objective}
\end{equation}

The topological inverse mapping $\pi_{\varphi_{ini}}$~\cite{poulenard2018topological,bruel2020topology,depeng2024persistent} links a persistent pair $(b,d)$ to a parametric coordinates $\xi_b$ and $\xi_d$:
\begin{equation}
    \left\{
        \begin{aligned}
            &\xi_b = (u_b,v_b,w_b) = \pi_{\varphi_{ini}}(b)\\
            &\xi_d = (u_d,v_d,w_d) = \pi_{\varphi_{ini}}(d),
        \end{aligned}
        \right.
    \label{eq: topological inverse mapping}    
\end{equation}    
The optimization aims to identify a blending function that avoids the topological errors of the blended porous structure within the blending regions while maintaining the structure beyond these regions.
    In this study, only persistent pairs relevant to the blending regions $BR$ are considered for optimization.
    To achieve this, a new set of persistent pairs $(\tilde{b},\tilde{d})$ is introduced as follows:
\begin{equation}
    (\tilde{b},\tilde{d}) = 
    \begin{cases}
        (b,d) & \xi_b \in BR ~\text{and} ~\xi_d \in BR \\
        (0,0) & \text{others}.
    \end{cases} 
    \label{eq: new persistent pair}
\end{equation}
    Finally, the new topological objective function is defined as follows:
\begin{equation}
    \tilde{L} = \sum_{k=0,2}\left( \sum_{i=0}^{N_k^\text{III}-1}\tilde{d_{i,k}^\text{III}} -
    \sum_{i=0}^{N_k^\text{II}-1}\tilde{b_{i,k}^\text{II}}\right).
    \label{eq: new optimization objective}
\end{equation}

The gradient of $\tilde{L}$ with respect to control coefficients $C_i$ is computed utilizing the topological inverse mapping $\pi_{\varphi_{ini}}$~\cite{depeng2024persistent}.
    The iterative adjustment of control coefficients employs the adaptive gradient descent method~\cite{duchi2011adaptive}, terminating the process upon reaching $\tilde{L} = 0$ or the maximum iteration limit.

\subsection{Blending scheme for more than two porous structures}
\label{subsec: 3.3 Blending scheme for more than two porous structures}
The merging of multiple porous structures can be accomplished through an iterative blending of two porous structures.

The initialization process described in Subsection~\ref{subsec: 3.1 Initialization of blending function} and the optimization method outlined in Subsection~\ref{subsec: 3.2 Optimization of blending function} can be integrated into a function labeled "BLEND." 
    When provided with two porous structures $\{\varphi_0\leq c_0\}$, $\{\varphi_1\leq c_1\}$, $ER_0$, $ER_1$, and $BR$, the function "BLEND" yields the associated function $\varphi$ of the blended structure $\Phi$.

\begin{figure}[H]
\centering
\resizebox{0.8\textwidth}{!}{%
\begin{minipage}{\textwidth}
\begin{algorithm}[H]
    \renewcommand{\algorithmicrequire}{\textbf{Input:}}
	\renewcommand{\algorithmicensure}{\textbf{Output:}}
    \caption{BLEND}
    \label{algorithm: BLEND}
\begin{algorithmic}[1]
\REQUIRE $\varphi_0$, $c_0$, $\varphi_1$, $c_1$, $ER_0$, $ER_1$, and $BR$
\STATE Initialize blending function $\omega(x,y,z)$ with method in Subsection~\ref{subsec: 3.1 Initialization of blending function}
\STATE Optimize blending function $\omega(x,y,z)$ with method in Subsection~\ref{subsec: 3.2 Optimization of blending function}  
\STATE $\varphi = (1-\omega)(\varphi_0-c_0)+\omega(\varphi_1-c_1)$; $\Phi = \{(x,y,z)~|~\varphi(x,y,z)\leq 0\}$
\ENSURE $\varphi$
\end{algorithmic}
\end{algorithm}
\end{minipage}%
}
\end{figure}    

Given a collection of porous structures $\{\varphi_i\leq c_i\}_{i=0}^N$ and their corresponding existing region sets $\{ER_i\}_{i=0}^N$, these porous structures undergo pairwise merging. 
    Assuming the relevant blending regions are designated as $\{BR_i\}_{i=0}^{N-1}$, at the $k$-th stage of the merging procedure, the current existing region is denoted as $CER$, and the current porous structure corresponds to the function $\varphi$. 
    Consequently, following the guidance from algorithm~\ref{algorithm: Blending of Multiple Porous Structures}, porous structures can be successively blended to yield the ultimate blended porous structure.

\begin{figure}[H]
\centering
\resizebox{0.8\textwidth}{!}{%
\begin{minipage}{\textwidth}
\begin{algorithm}[H]
    \renewcommand{\algorithmicrequire}{\textbf{Input:}}
	\renewcommand{\algorithmicensure}{\textbf{Output:}}
    \caption{Blending of Multiple Porous Structures}
    \label{algorithm: Blending of Multiple Porous Structures}
\begin{algorithmic}[1]
\REQUIRE $\{\varphi_i\leq c_i\}_{i=0}^N$, $\{ER_i\}_{i=0}^N$, $\{BR_i\}_{i=0}^{N-1}$
\STATE $\varphi \gets \varphi_0-c_0$; $CER \gets ER_0$
\FOR{$i = 0$ \textbf{to} $N-1$}
\STATE $\varphi$ $\gets$ BLEND($\varphi$,$0$,$\varphi_{i+1}$,$c_{i+1}$,$CER$,$ER_{i+1}$,$BR_i$)
\STATE $CER \gets CER \cup ER_{i+1}$
\ENDFOR
\STATE $\Phi = \{(x,y,z)~|~\varphi(x,y,z)\leq 0\}$
\ENSURE $\varphi$ and $\Phi$ 
\end{algorithmic}
\end{algorithm}
\end{minipage}%
}
\end{figure}   

\section{Experiments and discussions}
\label{sec: Experiments and discussions}
The design method proposed in this study is implemented using the C++ programming language and evaluated on a personal computer featuring an i7-10700 CPU running at 2.90 GHz with 16 GB of RAM.
    This section provides the experiments aimed at validating the efficacy of the proposed approach. 
   
In our experiments, regions are represented by implicit functions facilitating the location determination. 
    Specifically, the implicit function corresponding to the existing region $ER$ is defined as:
\begin{equation}
    \begin{cases}
        f_{ER}(x,y,z) \leq 0 & \text{(x, y, z)} \in  ER \\
        f_{ER}(x,y,z) > 0 & \text{others}.
    \end{cases}
\end{equation}
The detailed parameter settings for the experiments in this section can refer to Table~\ref{tab: information record}.

\subsection{One-dimensional blending situations}
\label{subsec: Simple blending situations}

\subsubsection{Blending of porous structures}
\label{subsubsec: Blending of porous structures}
\begin{figure}[h]
    \centering
        \includegraphics[width = 0.98\textwidth]{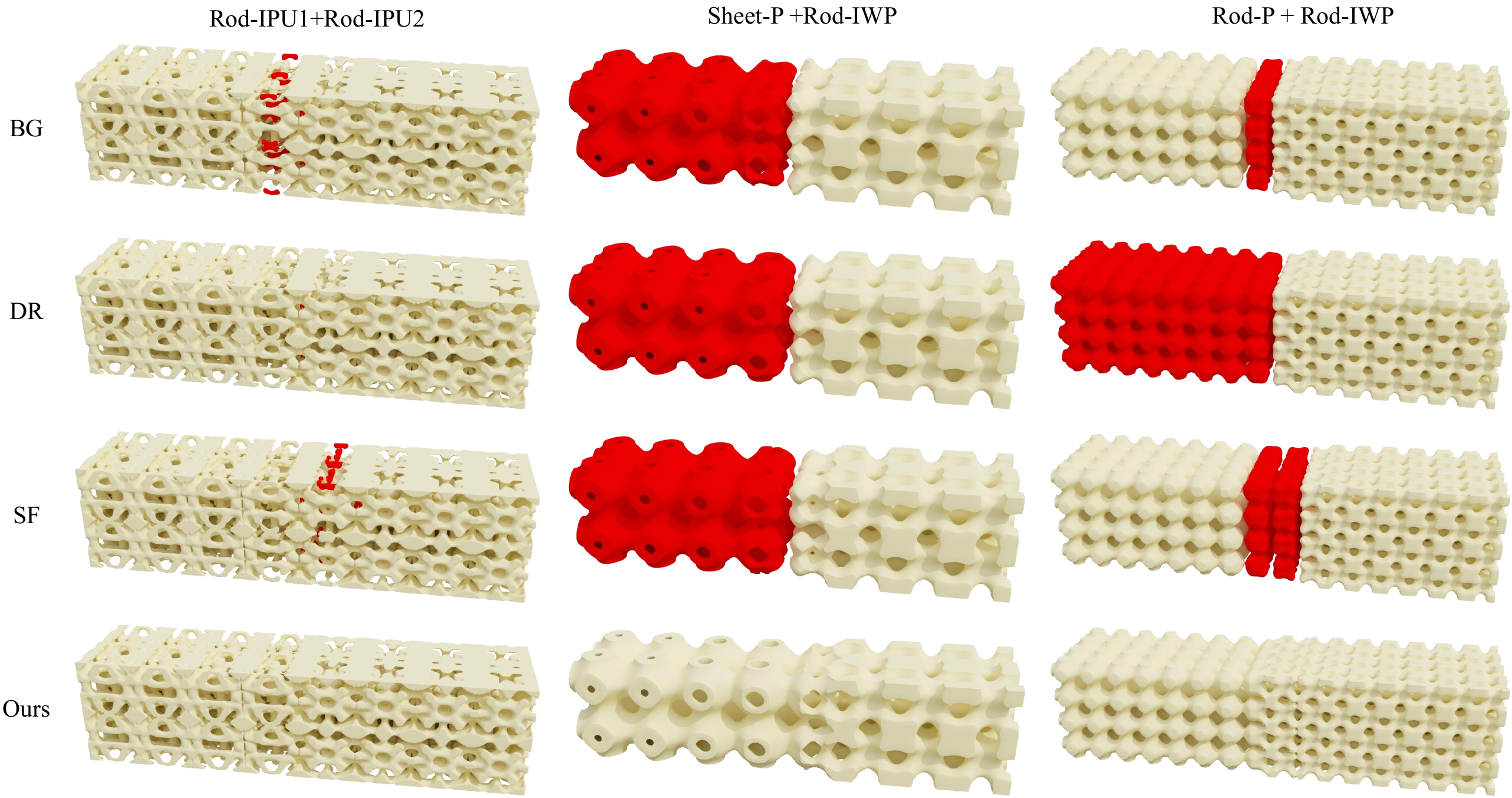}    
        \caption{Comparison of different blending methods. Isolated components are marked in red. The first row shows the results of Beta Growth method in~\cite{yoo2015advanced}. The second row shows the results of discrete reconstruction method in~\cite{ozdemir2023novel}. The third row shows the results of Sigmoid function method in~\cite{yang2014multi}. The last row shows the results of our method.}
        \label{fig: compare_connected_component}
\end{figure}

In this section, various types of TPMSs and implicit porous structures represented by periodic B-splines, as proposed in reference~\cite{gao2024periodic}, are employed to validate the efficacy of the method.
    Figure~\ref{fig: compare_connected_component} showcases the blending of Rod-IWP, Rod-P, Sheet-P, Sheet-IWP, and implicit periodic units (Rod-IPU). 
    The porous structure on the left side is located at $ER_1 = [0,0.5]\times[0,0.25]\times[0,0.25]$, while the porous structure on the right side is located at $ER_2 = [0.5,1.0]\times[0,0.25]\times[0,0.25]$.
    Various blending techniques including the Sigmoid function (SF) method from~\cite{yang2014multi}, the Beta Growth (BG) method from~\cite{yoo2015advanced}, and the discrete reconstruction (DR) from~\cite{ozdemir2023novel} are contrasted with our approach in the experiments. 
    In all experiments, the blending region $BR$ is set as $[0.3,0.7]\times[0,0.25]\times[0,0.25]$ via parameter adjustments.
    
\begin{figure}[h]
    \centering
        \includegraphics[width = 0.98\textwidth]{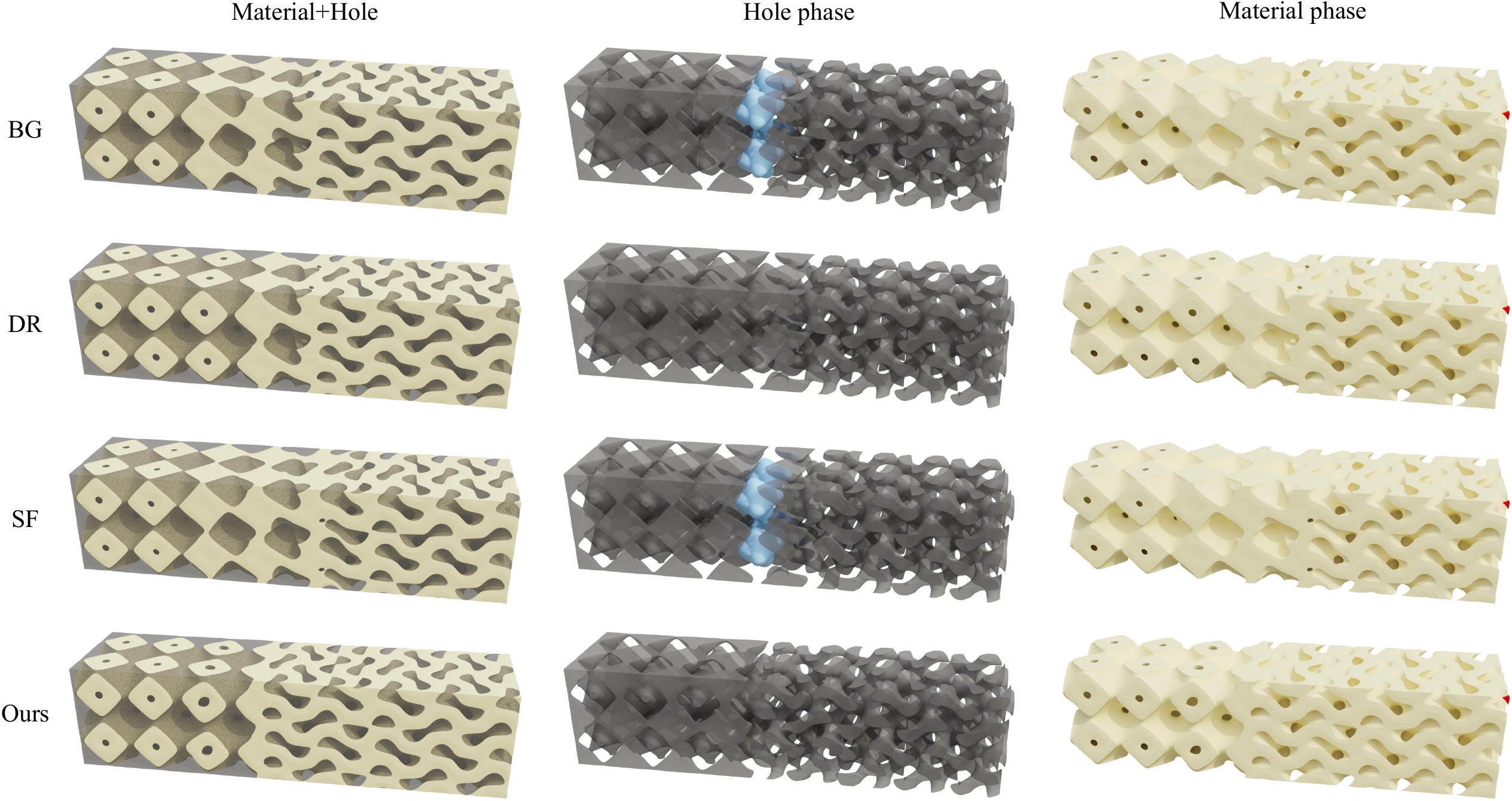}    
        \caption{Comparison of different blending methods. Isolated holes are marked in blue. Isolated connected components are marked in red. The first row shows the results of the Beta Growth method in~\cite{yoo2015advanced}. The second row shows the results of the discrete reconstruction method in~\cite{ozdemir2023novel}. The third row shows the results of the Sigmoid function method in~\cite{yang2014multi}. The last row shows the results of our method.}
        \label{fig: compare_isolated_holes}
\end{figure}

The blending results depicted in Figure~\ref{fig: compare_connected_component} highlight isolated connected components labeled in red, the number of isolated holes and connected components is recorded in Table~\ref{tab: Comparison} at the same time.
    Notably, our method stands out by consistently avoiding the generation of isolated connected components across three blending scenarios.
    Furthermore, Figure~\ref{fig: compare_isolated_holes} showcases isolated holes identified in blue within the hole phase, providing a visual comparison with the blending outcomes.
    The successful blending of Sheet-P and Sheet-G demonstrates our method's effectiveness in preventing the occurrence of isolated holes. 
    The final row of Table~\ref{tab: Comparison} shows that our method and the DR method effectively avoid isolated holes.
    Moreover, our blending strategy achieves a more seamless and natural transition, as evident from visual inspections. 
    It is crucial to note, as visualized in Figure~\ref{fig: compare_isolated_holes}, the presence of an intrinsic isolated connected component in the upper right corner of the blended structures, unaffected by the blending method.
    The correct count of connected components in the final result should remain at $2$.

\begin{table}[]
\centering
\caption{The number of topological features of the blended porous structures in Figure~\ref{fig: compare_connected_component} and Figure~\ref{fig: compare_isolated_holes}. $n_0$ represents the number of connected components. $n_2$ represents the number of isolated holes. The blended porous structures without topological errors are highlighted.}
\label{tab: Comparison}
\resizebox{0.8\linewidth}{!}{
\begin{tabular}{lcclcclcclcc}
\hline
\multicolumn{1}{c}{} & \multicolumn{2}{c}{BG} &  & \multicolumn{2}{c}{DR}  &  & \multicolumn{2}{c}{SF} &  & \multicolumn{2}{c}{Ours} \\ \cline{2-3} \cline{5-6} \cline{8-9} \cline{11-12} 
\multicolumn{1}{c}{} & $n_0$      & $n_2$     &  & $n_0$      & $n_2$      &  & $n_0$      & $n_2$     &  & $n_0$       & $n_2$      \\ \hline
Rod-IPU1 + Rod-IPU2  & 30         & 4         &  & 9          & 4          &  & 34         & 4         &  & \textbf{1}  & \textbf{0} \\
Rod-P + Sheet-P      & 2          & 0         &  & 2          & 0          &  & 2          & 0         &  & \textbf{1}  & \textbf{0} \\
Rod-P + Rod-IWP      & 18         & 0         &  & 2          & 0          &  & 34         & 0         &  & \textbf{1}  & \textbf{0} \\
Sheet-P + Sheet-G    & 2          & 9         &  & \textbf{2} & \textbf{0} &  & 2          & 6         &  & \textbf{2}  & \textbf{0} \\ \hline
\end{tabular}}
\end{table}

\subsubsection{Blending of free-form models}
\label{subsubsec: Blending of free-form models}
As stated in the work by Yoo~\cite{yoo2011porous}, when considering a model $M$ represented by an implicit function $M={\phi_M \leq 0}$, the porous model $\tilde{M}$ containing a porous structure defined by $\{\varphi \leq 0\}$ is formulated as follows:
\begin{equation}
    \left\{
        \begin{aligned}
            &\tilde{M} = \{(x,y,z)~|~\tilde{\varphi}(x,y,z)\leq 0\}\\
            &\tilde{\varphi} = \max(\varphi,\phi_M).
        \end{aligned}
    \right.
\end{equation}  
    In our approach, utilizing $\tilde{\varphi}$ instead of $\varphi$ effectively handles the blending of free-form porous models.

\begin{figure}[h]
    \centering
        \includegraphics[width = 0.98\textwidth]{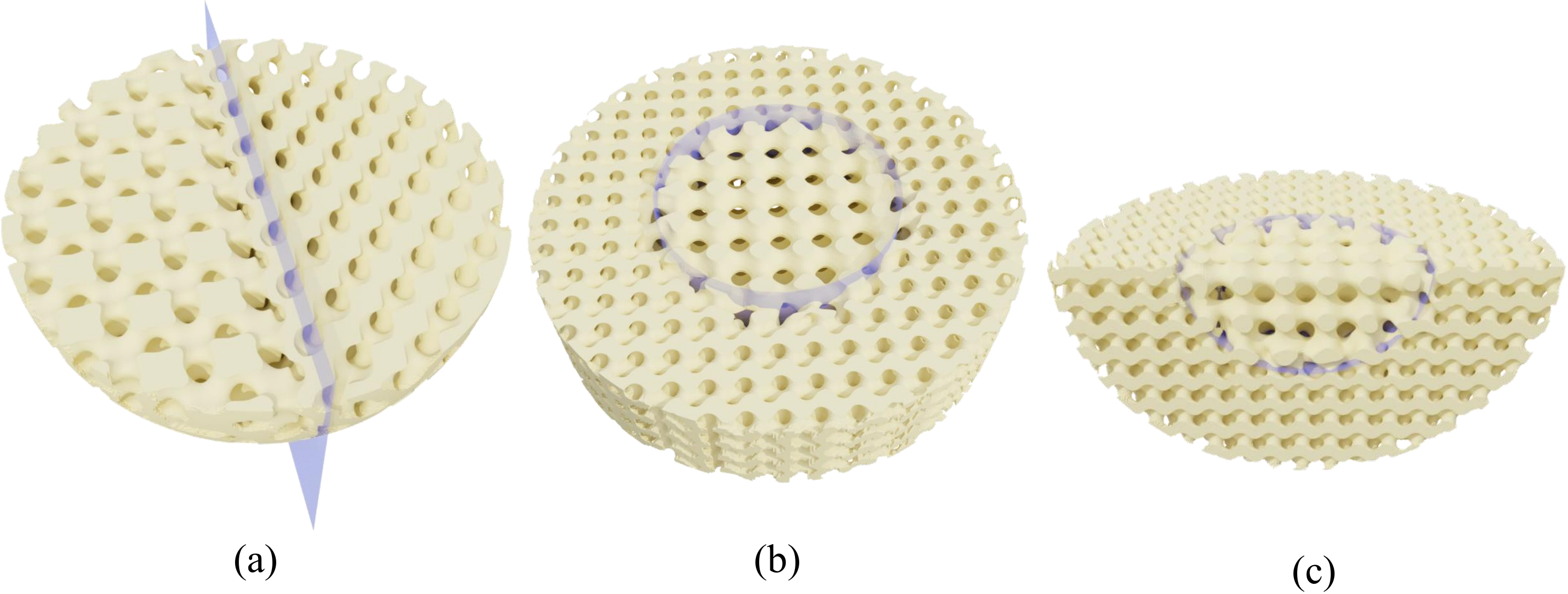}    
        \caption{Porous models are blended along the blue boundary. (a) The blending region is $0.4\leq x\leq 0.6$ (b) The blending region is $0.2\leq r\leq 0.3$. (c) The blending region is $0.15\leq r\leq 0.35$.}
        \label{fig: Blending of porous models}
\end{figure}

As shown in Figure~\ref{fig: Blending of porous models}, different porous models are situated on each side of the blue boundary. 
    The blending region in Figure~\ref{fig: Blending of porous models}(a) is defined as $0.4\leq x\leq 0.6$ using the one-dimensional blending scheme.
    As illustrated in Subsection~\ref{subsec: 3.1.1 One-dimensional blending situation}, the scenarios in Figure~\ref{fig: Blending of porous models}(b) and (c) can be viewed as blending along the axis direction. 
    Therefore, the blending region in Figure~\ref{fig: Blending of porous models}(b) is defined as $0.2\leq r\leq 0.3$, where $r$ represents the radius of the cylinder section. 
    Additionally, the blending region in Figure~\ref{fig: Blending of porous models}(c) is specified as $0.15\leq r\leq 0.35$, where $r$ represents the radius of the quarter sphere. 
    These three blended models in Figure~\ref{fig: Blending of porous models} exhibit no additional isolated connected components or isolated holes, indicating the efficacy of our method in the blending of porous models.

\subsection{Three-dimensional blending situations}
\label{subsec: Complex blending situations}
\begin{figure}[h]
    \centering
        \includegraphics[width = 0.98\textwidth]{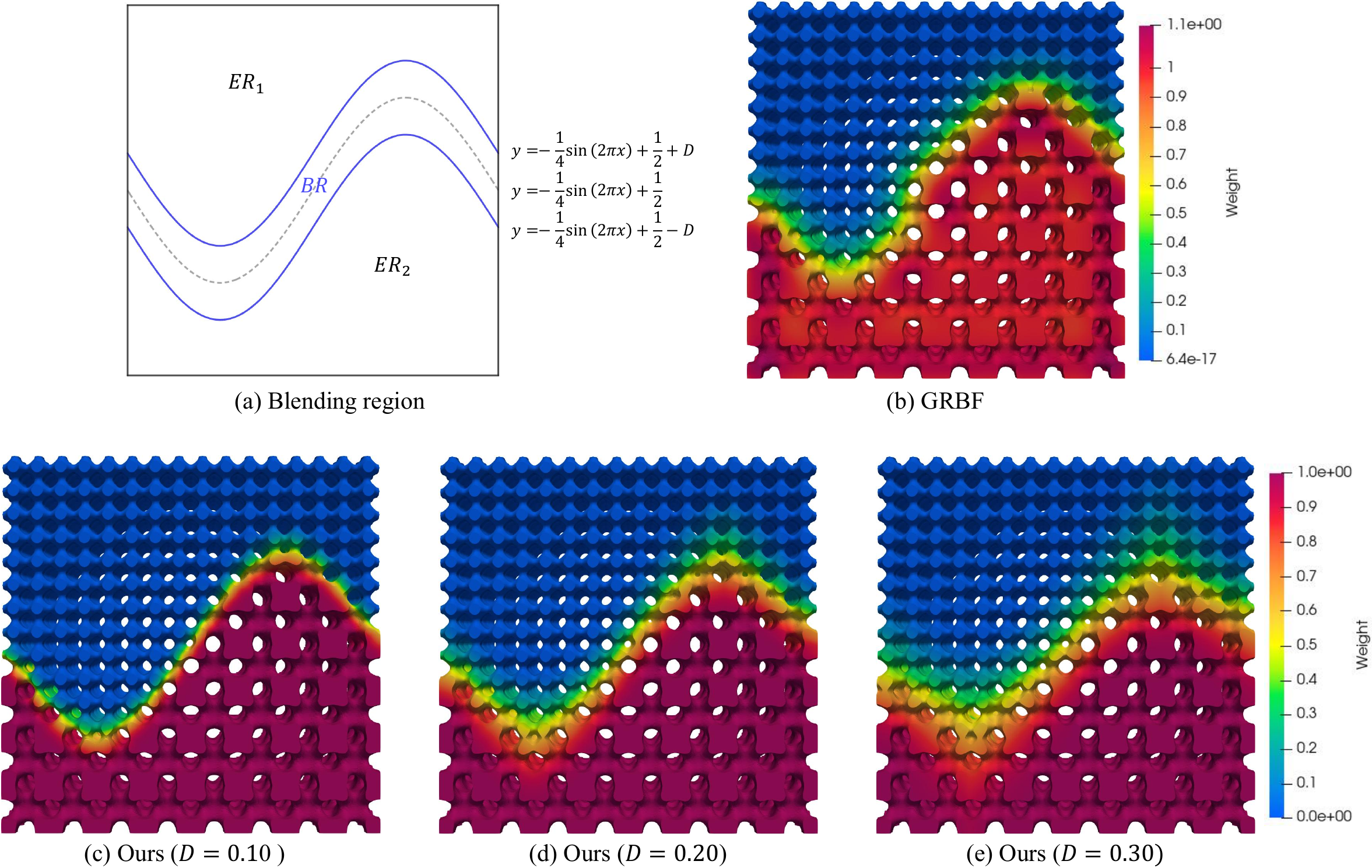}    
        \caption{Blending experiment under complex blending region. The color of the blended structure represents the weight function value at the corresponding position. (a) Description of the blending region. (b) The result of the GRBF method. (c) The result of our method under the blending region corresponds to $D=0.1$. (d) The result of our method under the blending region corresponds to $D=0.2$. (e) The result of our method under the blending region corresponds to $D=0.3$.}
        \label{fig: sinx_blending}
\end{figure}

In this subsection, initially, two porous structures are blended along the complex blending region as shown in Figure~\ref{fig: sinx_blending}(a). 
    The one-dimensional blending method, SF method, BG method, and DR method discussed in Subsection~\ref{subsec: Simple blending situations} are inadequate for such intricate scenarios. 
    Yang et al.~\cite{yang2014multi} addressed these challenges by employing Gaussian Radial Basis Functions (GRBF) as weight functions.
    The fundamental concept of the GRBF method involves sampling numerous points within $ER_1$ and $ER_2$, assigning them values of $0$ and $1$ as data points, respectively. 
    Subsequently, Gaussian Radial Basis Functions are defined on these data points to interpolate them, generating a weight function. 
    We sampled 20 data points in each direction and specified the parameter $\delta$ of the radial basis function as $0.02$. 
    The outcomes are presented in Figure~\ref{fig: sinx_blending}(b), where the color of the blended structure corresponds to the weight at the respective position.
    Although the blended shapes from the GRBF method align with expectations, it does not ensure that structures outside the blending region remain unchanged and lack precise specification of the blending region.

Our method offers precise control over the shape of the blending region.
    In this experiment, the boundary surface is shifted along the $y$-direction by $D$ to generate the blended region. 
    When the offset of $D$ is 0.1, 0.2, and 0.3 respectively, the resulting blended structures are shown in Figure~\ref{fig: sinx_blending}(a), (b), and (c). 
    During initialization, control coefficients outside the blending region are set to 0 and 1.  
    Subsequently, as the data points are sampled within the blending region during the fitting stage, due to the local support property of B-spline basis values beyond this region remain unaffected.
    Only the control coefficients whose support domain is within the blending region are updated during optimization, ensuring the preservation of the structure outside the blending region.   
    Additionally, this experiment demonstrates that our method can freely and accurately control the shape of the blending region.

\begin{figure}[h]
    \centering
        \includegraphics[width = 0.98\textwidth]{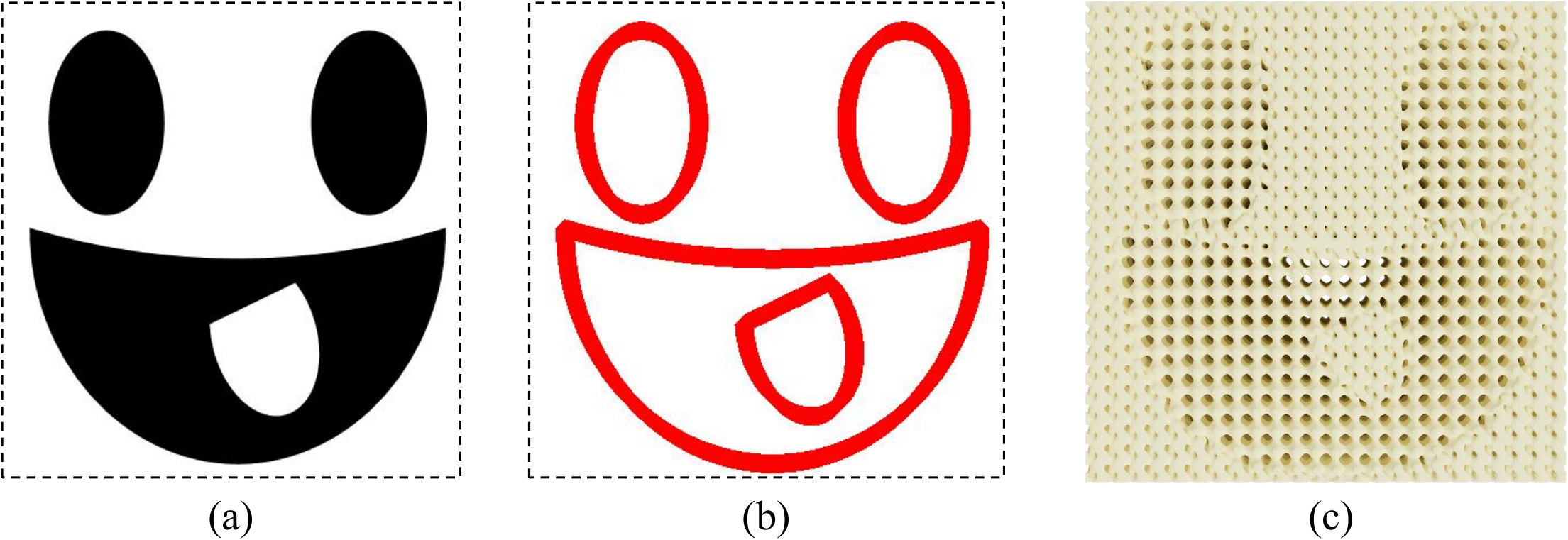}    
        \caption{(a) Region configuration is defined by an image. Rod-P TPMS is located in the black region, and Rod-G TPMS is located in the white region. (b) Blending region. (c) Blended porous structure.}
        \label{fig: smile_image}
\end{figure}

To showcase the method's effectiveness within disconnected blending regions, we construct an image as shown in Figure~\ref{fig: smile_image}(a) for region configuration.
    The black and white regions consist of Rod-P and Rod-G TPMSs, respectively. 
    By expanding the boundary pixels inward and outward by 8 pixels, the obtained blending region is displayed in Figure~\ref{fig: smile_image}(b).
    The blended porous structure aligns well with the defined regions in the image and the intended blending region, avoiding isolated holes or additional connected components.
    Our method initializes the weight function by calculating the distance from the data points to the blending boundary.  
    Therefore, even with a blending region composed of four connected components, our method achieves satisfactory results.

\subsection{Discussion}
\label{subsec: discussion}
The experiments in this section show that our method works well in handling blending problems in one-dimensional and three-dimensional situations. 
    Compared to other methods, the main advantage of our method lies in considering topological issues, thereby effectively avoiding topological errors. 
    However, this also means that our method requires more computational time.

As shown in Table~\ref{tab: information record}, the calculation time is not significantly affected by the number of control coefficients. 
    This is because, on the one hand, the computational load involved in control coefficient iteration is small. 
    On the other hand, B-spline functions have local support property, with most control coefficients not participating in the iteration. 
    Since the topological features of the optimized porous structure need to be recalculated at each iteration, the number of optimization iterations and the discretization resolution when computing topological features are the main factors influencing the runtime. 
    To demonstrate this point, in the experiments in Figure~\ref{fig: sinx_blending}(c), we chose a resolution of $100\times 100\times 25$. 
    Although a blended porous structure with no topological errors is achieved after only $2$ iterations, a high resolution leads to significantly longer computation time compared with the experiment in Figure~\ref{fig: smile_image}. 

In our future work, we will focus on accelerating runtime. 
    On one hand, during the iteration process, changes occur only in local porous structures. The information of cubical complexes from the previous iterations does not need to be entirely recomputed. 
    We will utilize this feature to reduce the computation time for topological features.
    Additionally, as in the experiment in Figure~\ref{fig: sinx_blending}(c), resolutions of $100\times 100\times 25$ and $50\times 50\times 25$ both yield topologically correct structures. 
    Automatically selecting an appropriate resolution is also crucial for reducing computation time.

\begin{table}[]
\centering
\caption{Parameter settings for the experiments in Section~\ref{sec: Experiments and discussions}}
\label{tab: information record}
\resizebox{0.95\linewidth}{!}{
\begin{threeparttable}
\begin{tabular}{cccccc}
\hline
\multirow{2}{*}{Experiment}           & \multicolumn{2}{c}{Weight function}          & \multirow{2}{*}{Resolution\tnote{1}} & \multirow{2}{*}{Iterations\tnote{2}} & \multirow{2}{*}{Time (s)\tnote{3}} \\ \cline{2-3}
                                       & Control coefficient    & Degree              &                             &                             &                           \\ \hline
Rod-IPU1+Rod-IPU2                      & 50                     & 3                   & $50\times 50\times 50$      & 3                           & 3.34                      \\
Sheet-P+Rod-IWP                        & 50                     & 3                   & $50\times 50\times 50$      & 1                           & 1.43                      \\
Rod-P+Rod-IWP                          & 50                     & 3                   & $50\times 50\times 50$      & 1                           & 1.61                      \\
Figure~\ref{fig: compare_isolated_holes}    & 50                     & 3                   & $50\times 50\times 50$      & 1                           & 1.39                      \\
Figure~\ref{fig: Blending of porous models}(a) & 30                     & 3                   & $50\times 50\times 50$      & 1                           & 1.73                      \\
Figure~\ref{fig: Blending of porous models}(b) & 30                     & 3                   & $50\times 50\times 50$      & 4                           & 4.45                      \\
Figure~\ref{fig: Blending of porous models}(c) & 30                     & 3                   & $50\times 50\times 50$      & 9                           & 9.69                      \\
Figure~\ref{fig: sinx_blending}(c)    & $80\times 80\times 20$ & $3\times 3\times 3$ & $100\times 100\times 25$      & 2                           & 12.12                      \\
Figure~\ref{fig: smile_image} & $80\times 80\times 20$ & $3\times 3\times 3$ & $50\times 50\times 25$   & 3                           & 5.06                     \\ \hline
\end{tabular}
\begin{tablenotes}
\footnotesize
\item[1] Spatial discretization resolution when constructing cubical complexes.
\item[2] Number of iterations in the optimization process.
\item[3] Total time for generating blended porous structure.
\end{tablenotes}
\end{threeparttable}}
\end{table}

\section{Applications of the blending scheme}
\label{sec: Applications of the blending scheme}
In this section, some applications are demonstrated to showcase the potential applications of the blending method. 

\subsection{Porous scaffold design}
\label{subsec: Porous scaffold design}
The bone structure comprises a non-uniform porous configuration with density variations across spatial positions.
    Porous scaffolds attempt to mimic real bones in structural characteristics, Young's modulus, compression strength, biocompatibility, and other features~\cite{chen2020porous}. 
    A well-designed porous scaffold holds potential applications in bone repair, bone research, and other related fields.

\begin{figure}[h]
    \centering
        \includegraphics[width = 0.8\textwidth]{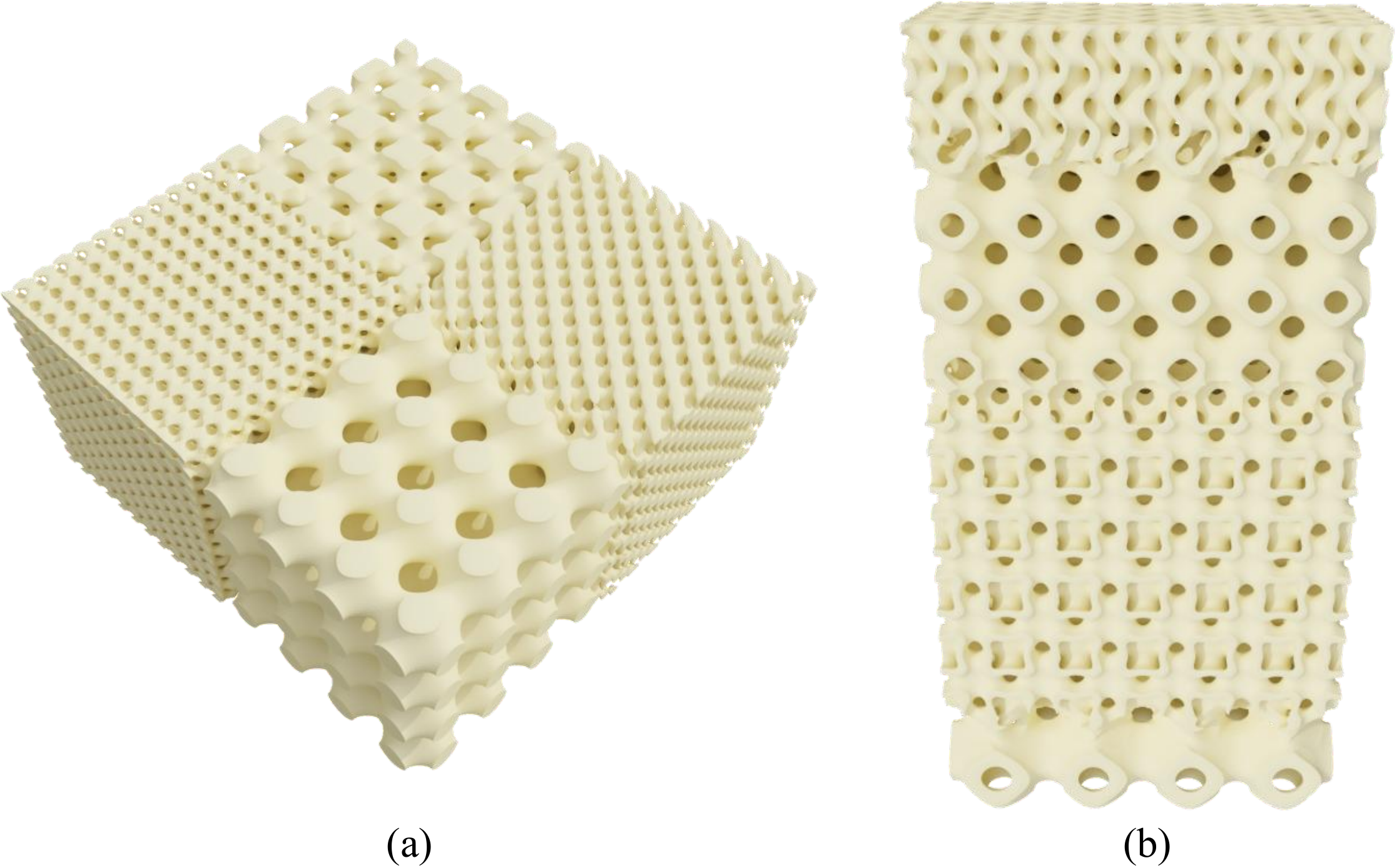}    
        \caption{(a) A porous scaffold consists of four different porous structures. (b) A porous scaffold composed of four different porous structures along the $z$-direction.}
        \label{fig: Application_TPMSs_Scaffold}
\end{figure}

Utilizing a porous scaffold featuring distinct biomimetic regions resembling the primary cancer tissue site and potential metastatic sites can facilitate the investigation of metastatic mechanisms, thereby aiding in the discovery of pharmaceutical interventions for metastasis prevention~\cite{vijayavenkataraman2018triply}. 
    To generate such a porous scaffold, four regions are designated utilizing Rod-P, Rod-G, Rod-D, and Rod-IWP TPMSs with varying pore sizes and densities. 
    The blending algorithm introduced in Subsection~\ref{subsec: 3.3 Blending scheme for more than two porous structures} is utilized to blend these four porous structures, resulting in the porous scaffold depicted in Figure~\ref{fig: Application_TPMSs_Scaffold}(a). 
    This scaffold exhibits a natural blended region without topological errors.

Natural bones exhibit spatial structural and functional gradients~\cite{vijayavenkataraman2018triply}. 
    Our approach is capable of producing a biomimetic gradient porous scaffold.
    As illustrated in Figure~\ref{fig: Application_TPMSs_Scaffold}(b), a porous scaffold blended from four porous structures with different shapes along the z-direction is generated. 
    By designing porous structures across distinct zones, our method proves effective in crafting biocompatible scaffolds without topological errors.

\begin{figure}[h]
    \centering
        \includegraphics[width = 0.7\textwidth]{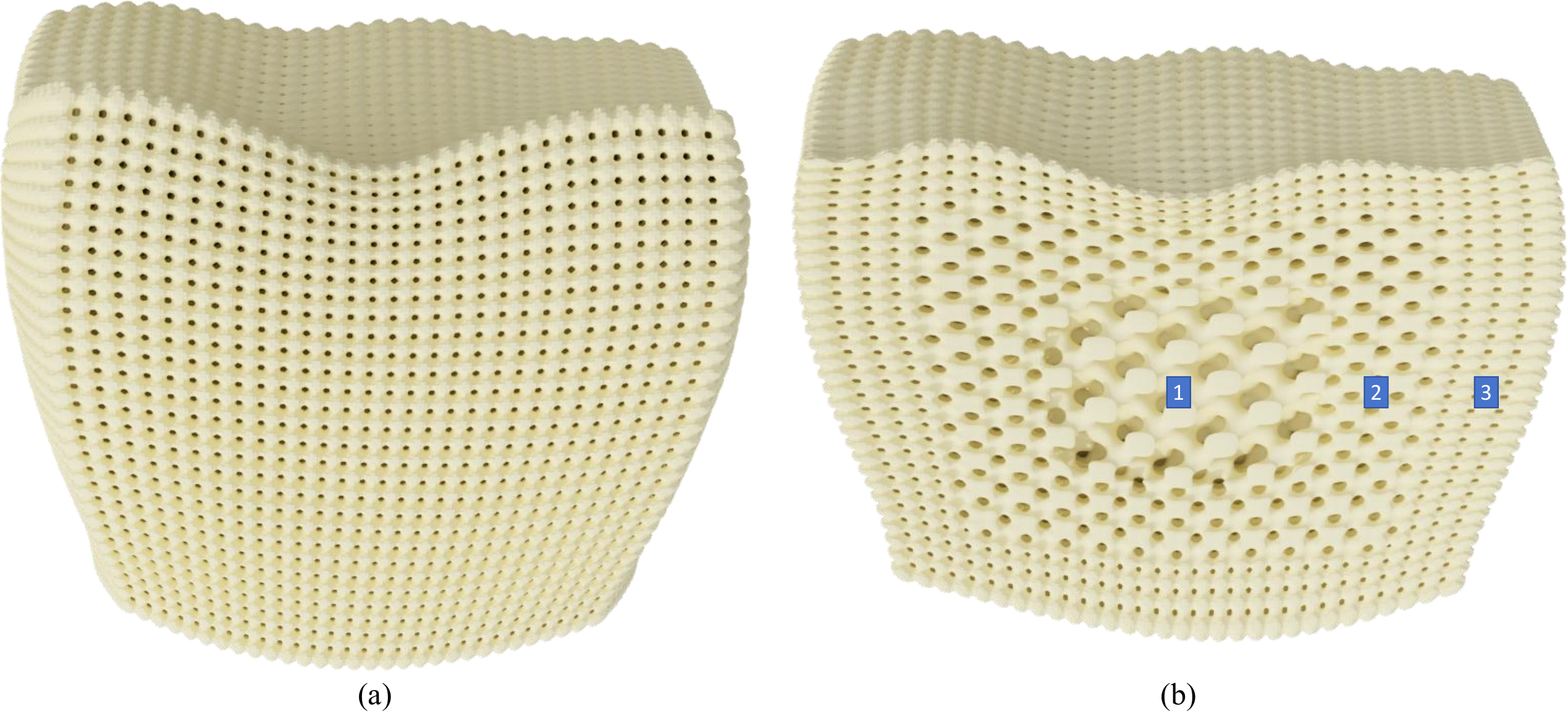}    
        \caption{(a) A tooth model. (b) Sectional view of the tooth model. There are three types of porous structures in regions 1, 2, and 3, with the relative density increasing progressively.}
        \label{fig: teeth_model}
\end{figure}

The proposed method allows for the design of more intricate porous scaffolds.
    As proposed in~\cite{hu2021heterogeneous}, porous models are initially constructed within a standardized design domain (unit cube) and then mapped to a free-form model. 
    To mimic bone structure, we generate a porous scaffold within a regular design domain with gradually increasing relative density from the interior to the exterior.
    The resulting scaffold is then mapped to the physical domain to produce the porous tooth model (detailed mapping techniques can be found in references~\cite{hu2021heterogeneous,gao2022connectivity}). 
    The porous scaffold and a sectional view are depicted in Figure~\ref{fig: teeth_model}, demonstrating our method's capability to blend multiple porous structures into a complex model.

\subsection{Multi-unit topology optimization}
\label{subsec: Multi-unit topology optimization}
An essential application of the proposed blending method lies in multi-unit topology optimization. 
    In conventional topology optimization for the porous model, the objective is to determine the optimal density distribution of the porous model given specific porous structures and boundary conditions~\cite{li2019design}. 
    Since different porous structures exhibit distinct mechanical properties, the multi-unit topology optimization approach seeks to optimize both the distribution of porous structure types and the density distribution within a set of porous structures~\cite{xu2023topology,feng2022stiffness}. 
    Once the structure types and density distribution are obtained, a blending method becomes necessary to blend various porous structures into a unified model.
    
\begin{figure}[h]
    \centering
        \includegraphics[width = 0.98\textwidth]{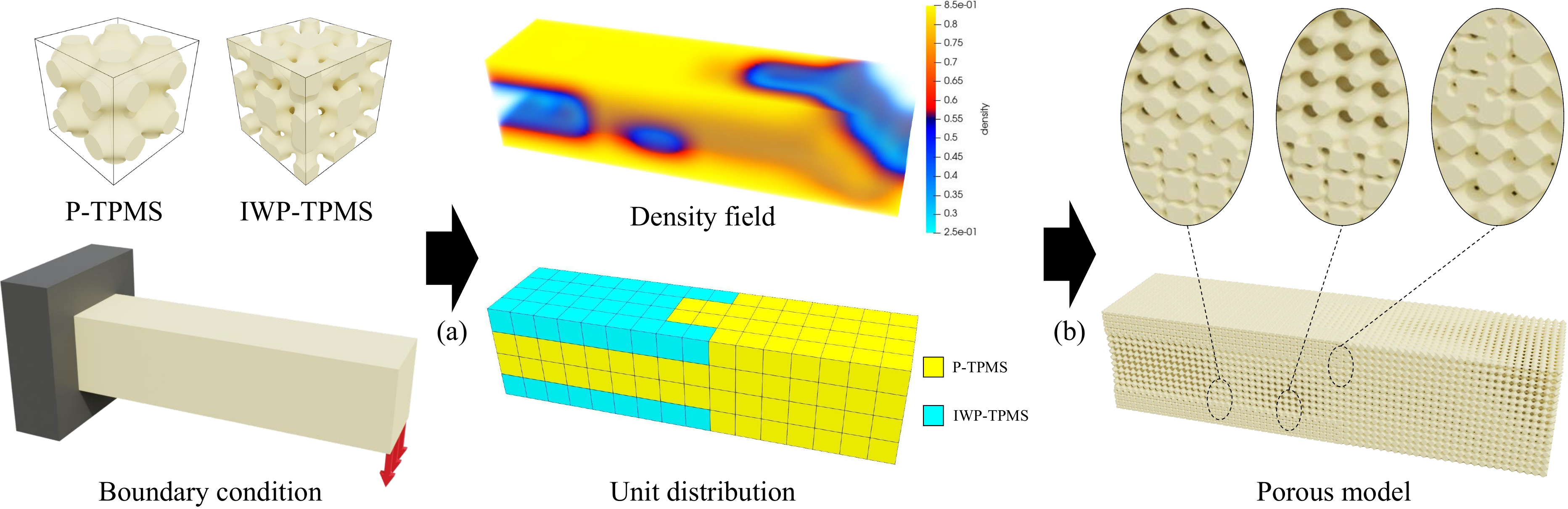}    
        \caption{Multi-unit topology optimization. (a) Determining the optimal unit distribution and density field using the method in~\cite{feng2022stiffness}. (b) Generating porous structures conforming to the density field using the method in~\cite{hu2021heterogeneous}, and then blending them using our method.}
        \label{fig: multi_unit_to}
\end{figure}

In Figure~\ref{fig: multi_unit_to}, the boundary conditions are illustrated, with one end of the beam fixed and the opposite end subjected to a uniform downward force. The volume fraction of the porous model is set to 0.5.
    Employing the topology optimization technique detailed in reference~\cite{feng2022stiffness}, the optimal density distribution and unit distribution (refer to the second column in Figure~\ref{fig: multi_unit_to}) are acquired.
    Following this, the density control method presented in~\cite{hu2021heterogeneous} is applied to achieve a porous model that aligns with the specified density field. 
    Subsequently, our proposed blending method is employed to merge two distinct porous structures, resulting in the final porous model.
    The results show that our method produces natural and topologically correct blending structures at the interfaces of different types of porous structures.
    Hence, our method exhibits promise as a post-processing approach for multi-unit topology optimization outputs.

\section{Conclusion}
\label{sec: Conclusion}
In this study, we introduce a blending method for designing implicit heterogeneous porous models.
    Different initialization methods are proposed to address various blending scenarios, resulting in an initial blended porous structure.
    Subsequently, the elimination of topological errors in the blended porous structure is formulated as an optimization problem based on persistent homology.  
    The optimization procedure is solved using a gradient-based algorithm while maintaining the structure outside the blending region unchanged. 
    Experimental results demonstrate that the blended structure obtained through our blending method exhibits no topological errors (isolated connected components and isolated holes) compared to alternative methods.
    Moreover, the proposed blending method is applied to design porous scaffolds and high-stiffness heterogeneous porous models, showcasing the versatility and potential applications of this approach.

Our blending method primarily focuses on addressing topological errors to define the optimization problem.
    Future directions may involve integrating additional geometric properties like mean curvature, density, and pore size into the optimization objective.
    Section~\ref{sec: Applications of the blending scheme} illustrates the potential applications of our proposed blending method. 
    Subsequent studies should aim to design porous scaffolds that meet practical application requirements like porosity and pore size more effectively. 
    Furthermore, the performance of heterogeneous models obtained through multi-unit topology optimization is intricately linked to the geometric morphology of the blending region. 
    Addressing this challenge necessitates further investigation into blending methodologies that minimize stress concentrations within the blended structure.

\section*{Acknowledgments}
\noindent
This work is supported by National Natural Science Foundation of China under Grant nos. 62272406 and 61932018.

\section*{Declaration of competing interest}
The authors declare that they have no known competing financial interests or personal relationships that could have appeared to influence the work reported in this paper.

\section*{Declaration of Generative AI and AI-assisted technologies in the writing process}
During the preparation of this work the author(s) used ChatGPT in order to improve language and readability. After using this tool/service, the author(s) reviewed and edited the content as needed and take(s) full responsibility for the content of the publication.

\bibliographystyle{elsarticle-num} 
\bibliography{mybibfile}

\begin{thebibliography}{10}
\expandafter\ifx\csname url\endcsname\relax
  \def\url#1{\texttt{#1}}\fi
\expandafter\ifx\csname urlprefix\endcsname\relax\def\urlprefix{URL }\fi
\expandafter\ifx\csname href\endcsname\relax
  \def\href#1#2{#2} \def\path#1{#1}\fi

\bibitem{chen2020porous}
H.~Chen, Q.~Han, C.~Wang, Y.~Liu, B.~Chen, J.~Wang, Porous scaffold design for
  additive manufacturing in orthopedics: a review, Frontiers in bioengineering
  and biotechnology 8 (2020) 609.

\bibitem{shi2021design}
X.~Shi, W.~Liao, T.~Liu, C.~Zhang, D.~Li, W.~Jiang, C.~Wang, F.~Ren, Design
  optimization of multimorphology surface-based lattice structures with density
  gradients, The International Journal of Advanced Manufacturing Technology 117
  (2021) 2013--2028.

\bibitem{feng2022stiffness}
Y.~Feng, T.~Huang, Y.~Gong, P.~Jia, Stiffness optimization design for tpms
  architected cellular materials, Materials \& Design 222 (2022) 111078.

\bibitem{xu2023topology}
W.~Xu, P.~Zhang, M.~Yu, L.~Yang, W.~Wang, L.~Liu, Topology optimization via
  spatially-varying tpms, IEEE Transactions on Visualization and Computer
  Graphics (2023).

\bibitem{yan2019strong}
X.~Yan, C.~Rao, L.~Lu, A.~Sharf, H.~Zhao, B.~Chen, Strong 3d printing by tpms
  injection, IEEE Transactions on Visualization and Computer Graphics 26~(10)
  (2019) 3037--3050.

\bibitem{gao2024periodic}
D.~Gao, Y.~Gao, H.~Lin, Periodic implicit representation, design and
  optimization of porous structures using periodic b-splines, Computer-Aided
  Design (2024) 103703.

\bibitem{hong2021conformal}
Q.~Y. Hong, G.~Elber, Conformal microstructure synthesis in trimmed trivariate
  based v-reps, Computer-Aided Design 140 (2021) 103085.

\bibitem{gourmel2013gradient}
O.~Gourmel, L.~Barthe, M.-P. Cani, B.~Wyvill, A.~Bernhardt, M.~Paulin,
  H.~Grasberger, A gradient-based implicit blend, ACM Transactions on Graphics
  (TOG) 32~(2) (2013) 1--12.

\bibitem{pasko1995function}
A.~Pasko, V.~Adzhiev, A.~Sourin, V.~Savchenko, Function representation in
  geometric modeling: concepts, implementation and applications, The visual
  computer 11 (1995) 429--446.

\bibitem{yang2014multi}
N.~Yang, Z.~Quan, D.~Zhang, Y.~Tian, Multi-morphology transition hybridization
  cad design of minimal surface porous structures for use in tissue
  engineering, Computer-Aided Design 56 (2014) 11--21.

\bibitem{yoo2015advanced}
D.-J. Yoo, K.-H. Kim, An advanced multi-morphology porous scaffold design
  method using volumetric distance field and beta growth function,
  International Journal of Precision Engineering and Manufacturing 16 (2015)
  2021--2032.

\bibitem{ren2021transition}
F.~Ren, C.~Zhang, W.~Liao, T.~Liu, D.~Li, X.~Shi, W.~Jiang, C.~Wang, J.~Qi,
  Y.~Chen, et~al., Transition boundaries and stiffness optimal design for
  multi-tpms lattices, Materials \& Design 210 (2021) 110062.

\bibitem{yoo2012heterogeneous}
D.-J. Yoo, Heterogeneous porous scaffold design for tissue engineering using
  triply periodic minimal surfaces, International Journal of Precision
  Engineering and Manufacturing 13 (2012) 527--537.

\bibitem{ozdemir2023novel}
M.~Ozdemir, U.~Simsek, G.~Kiziltas, C.~E. Gayir, A.~Celik, P.~Sendur, A novel
  design framework for generating functionally graded multi-morphology lattices
  via hybrid optimization and blending methods, Additive Manufacturing 70
  (2023) 103560.

\bibitem{zhang2023regulated}
X.~Zhang, L.~Jiang, X.~Yan, Z.~Wang, X.~Li, G.~Fang, Regulated multi-scale
  mechanical performance of functionally graded lattice materials based on
  multiple bioinspired patterns, Materials \& Design 226 (2023) 111564.

\bibitem{li2021simple}
Y.~Li, Q.~Xia, S.~Yoon, C.~Lee, B.~Lu, J.~Kim, Simple and efficient volume
  merging method for triply periodic minimal structures, Computer Physics
  Communications 264 (2021) 107956.

\bibitem{piegl2012nurbs}
L.~Piegl, W.~Tiller, The NURBS book, Springer Science \& Business Media, 2012.

\bibitem{feng2019efficient}
J.~Feng, J.~Fu, C.~Shang, Z.~Lin, X.~Niu, B.~Li, Efficient generation strategy
  for hierarchical porous scaffolds with freeform external geometries, addit.
  manuf. 31 (2020) 100943 (2019).

\bibitem{hu2021heterogeneous}
C.~Hu, H.~Lin, Heterogeneous porous scaffold generation using trivariate
  b-spline solids and triply periodic minimal surfaces, Graphical Models 115
  (2021) 101105.

\bibitem{yoo2011porous}
D.~J. Yoo, Porous scaffold design using the distance field and triply periodic
  minimal surface models, Biomaterials 32~(31) (2011) 7741--7754.

\bibitem{hong2023implicit}
Q.~Y. Hong, G.~Elber, M.-S. Kim, Implicit functionally graded conforming
  microstructures, Computer-Aided Design 162 (2023) 103548.

\bibitem{gao2022connectivity}
D.~Gao, J.~Chen, Z.~Dong, H.~Lin, Connectivity-guaranteed porous synthesis in
  free form model by persistent homology, Computers \& Graphics 106 (2022)
  33--44.

\bibitem{hu2023isogeometric}
C.~Hu, H.~Hu, H.~Lin, J.~Yan, Isogeometric analysis-based topological
  optimization for heterogeneous parametric porous structures, Journal of
  Systems Science and Complexity 36~(1) (2023) 29--52.

\bibitem{feng2021isotropic}
J.~Feng, B.~Liu, Z.~Lin, J.~Fu, Isotropic porous structure design methods based
  on triply periodic minimal surfaces, Materials \& Design 210 (2021) 110050.

\bibitem{vijayavenkataraman2018triply}
S.~Vijayavenkataraman, L.~Zhang, S.~Zhang, J.~Y. Hsi~Fuh, W.~F. Lu, Triply
  periodic minimal surfaces sheet scaffolds for tissue engineering
  applications: An optimization approach toward biomimetic scaffold design, ACS
  Applied Bio Materials 1~(2) (2018) 259--269.

\bibitem{local-lspia}
Y.~GAO, Y.~JIANG, H.~LIN, Local progressive and iterative approximation for
  least squares b-spline curve and surface fitting, Computer Science 51~(1)
  (2024) 225.

\bibitem{poulenard2018topological}
A.~Poulenard, P.~Skraba, M.~Ovsjanikov, Topological function optimization for
  continuous shape matching, in: Computer Graphics Forum, Vol.~37, Wiley Online
  Library, 2018, pp. 13--25.

\bibitem{bruel2020topology}
R.~Br{\"u}el-Gabrielsson, V.~Ganapathi-Subramanian, P.~Skraba, L.~J. Guibas,
  Topology-aware surface reconstruction for point clouds, in: Computer Graphics
  Forum, Vol.~39, Wiley Online Library, 2020, pp. 197--207.

\bibitem{depeng2024persistent}
G.~Depeng, Z.~Yuanzhi, L.~Hongwei, Persistent homology-driven optimization of
  effective relative density range for triply periodic minimal surface, arXiv
  preprint arXiv:2402.12109 (2024).

\bibitem{duchi2011adaptive}
J.~Duchi, E.~Hazan, Y.~Singer, Adaptive subgradient methods for online learning
  and stochastic optimization., Journal of machine learning research 12~(7)
  (2011).

\bibitem{li2019design}
D.~Li, N.~Dai, Y.~Tang, G.~Dong, Y.~F. Zhao, Design and optimization of graded
  cellular structures with triply periodic level surface-based topological
  shapes, Journal of Mechanical Design 141~(7) (2019) 071402.

\end{thebibliography}

\end{document}